\documentclass[10pt,journal]{IEEEtran}
\usepackage{multirow}
\usepackage{makecell}
\usepackage{amsmath,amsfonts}
\usepackage{array}
\usepackage[caption=false,font=normalsize,labelfont=sf,textfont=sf]{subfig}
\usepackage{textcomp}
\usepackage{stfloats}
\usepackage{url}
\usepackage{color}
\usepackage{verbatim}
\usepackage{graphicx}
\usepackage{cite}
\usepackage{amssymb}
\usepackage{booktabs}

\usepackage[linesnumbered,algoruled,boxed,lined]{algorithm2e}

\begin{document}
\title{
Encrypted Traffic Detection in Resource Constrained IoT Networks: A Diffusion Model and LLM Integrated Framework
}

\author{Hongjuan Li, 
        Hui Kang, 
        Chenbang Liu,
        Ruolin Wang,
        Jiahui Li,
        Geng Sun,~\IEEEmembership{Senior Member,~IEEE,}\\
        Jiacheng Wang,
        Shuang Liang,
        and Shiwen Mao,~\IEEEmembership{Fellow,~IEEE}
\thanks{
\par This work is supported in part by the National Key Research and Development Program of China (2024YFE03000104), in part by the National Natural Science Foundation of China (62272194, 62471200), in part by the Science and Technology Development Plan Project of Jilin Province (20240302079GX), in part by the Postdoctoral Fellowship Program of China Postdoctoral Science Foundation (GZC20240592), in part by China Postdoctoral Science Foundation General Fund (2024M761123), and in part by the Scientific Research Project of Jilin Provincial Department of Education (JJKH20250117KJ). \textit{(Corresponding author: Jiahui Li and Geng Sun.)}

\par Hongjuan Li, Hui Kang, Chenbang Liu, Roulin Wang, and Jiahui Li are with the College of Computer Science and Technology, Jilin University, Changchun 130012, China, and also with the Key Laboratory of Symbolic Computation and Knowledge Engineering of Ministry of Education, Jilin University, Changchun 130012, China (e-mails: hongjuan23@mails.jlu.edu.cn; kanghui@jlu.edu.cn; liucb23@mails.jlu.edu.cn; wangrl24@mails.jlu.edu.cn; lijiahui@jlu.edu.cn).

\par Geng Sun is with the College of Computer Science and Technology, Key Laboratory of Symbolic Computation and Knowledge Engineering of Ministry of Education, Jilin University, Changchun 130012, China, and also with the College of Computing and Data Science, Nanyang Technological University, Singapore 639798 (e-mail: sungeng@jlu.edu.cn).

\par Jiacheng Wang is with the College of Computing and Data Science, Nanyang Technological University, Singapore 639798 (e-mail: jiacheng.wang@ntu.edu.sg).

\par Shuang Liang is with the School of Information Science and Technology, Northeast Normal University, Changchun 130117, China (e-mail: liangshuang@nenu.edu.cn).

\par Shiwen Mao with the Department of Electrical and Computer Engineering, Auburn University, Auburn 36830, USA (e-mail: smao@ieee.org).
}
}



\IEEEtitleabstractindextext{
\begin{abstract}
The proliferation of Internet-of-things (IoT) infrastructures and the widespread adoption of traffic encryption present significant challenges, particularly in environments characterized by dynamic traffic patterns, constrained computational capabilities, and strict latency constraints. In this paper, we propose DMLITE, a diffusion model and large language model (LLM) integrated traffic embedding framework for network traffic detection within resource-limited IoT environments. The DMLITE overcomes these challenges through a tri-phase architecture including traffic visual preprocessing, diffusion-based multi-level feature extraction, and LLM-guided feature optimization. Specifically, the framework utilizes self-supervised diffusion models to capture both fine-grained and abstract patterns in encrypted traffic through multi-level feature fusion and contrastive learning with representative sample selection, thus enabling rapid adaptation to new traffic patterns with minimal labeled data. Furthermore, DMLITE incorporates LLMs to dynamically adjust particle swarm optimization parameters for intelligent feature selection by implementing a dual objective function that minimizes both classification error and variance across data distributions. Comprehensive experimental validation on benchmark datasets confirms the effectiveness of DMLITE, achieving classification accuracies of 98.87\%, 92.61\%, and 99.83\% on USTC-TFC, ISCX-VPN, and Edge-IIoTset datasets, respectively. This improves classification accuracy by an average of 3.7\% and reduces training time by an average of 41.9\% compared to the representative deep learning model.
\end{abstract}

}

\maketitle
\IEEEdisplaynontitleabstractindextext
\IEEEpeerreviewmaketitle

\section{Introduction}
\label{sec:introduction}
\par The rapid proliferation of Internet-of-things (IoT) technology is profoundly reshaping contemporary network ecosystems and digital infrastructure~\cite{TranDang2020, Li2025Aerial}. With connected IoT devices projected to exceed 75 billion by 2025~\cite{Alao2025, Li2024c}, network traffic detection has emerged as a critical component in ensuring the security and operational stability of these complex environments~\cite{Sheng2025}. Effective traffic detection not only identifies anomalous behaviors and malicious attacks but also supports optimized allocation of network resources and quality of service guarantees~\cite{Wu2021, Liu2024}. In increasingly sophisticated threat landscapes, particularly when facing zero-day attacks and encrypted traffic challenges~\cite{Stellios2018, Papadogiannaki2022}, precise traffic classification and anomaly detection mechanisms have become essential pillars for maintaining the integrity and reliability of IoT ecosystems~\cite{Dai2023}.

\par Despite significant advances in network traffic detection research~\cite{Nascita2024}, existing methods face mounting challenges as IoT environments become more complex and traffic encryption technologies become increasingly sophisticated and widespread~\cite{Lin2021}. Traditional traffic classification methods, including port-based approaches and deep packet inspection (DPI)~\cite{Blaise2020, Yan2019}, have become unsuitable for modern IoT network environments due to the prevalence of port disguising techniques and encrypted traffic~\cite{Zhu2023}. To address these limitations, traffic detection methods based on machine learning (ML)~\cite{Ede2020} and deep learning (DL) were subsequently introduced~\cite{Zhang2019, Zhang2025c}. ML methods aim to identify network traffic through manually designed features and trained classifiers~\cite{Gaurav2023}, offering improvements in pattern recognition capabilities. Concurrently, DL methods employ neural networks to automatically learn representations from raw traffic data, processing traffic in an end-to-end fashion by transforming it into image or sequence formats~\cite{Zhou2017, Xiao2022}.

\par However, both ML and DL approaches present significant limitations in IoT contexts~\cite{Xu2024, Zhang2024b}. ML-based methods such as ensemble classifiers and support vector machine (SVM) heavily rely on manually engineered features and struggle to adapt to dynamically changing network environments, thereby limiting their effectiveness against evolving threats and traffic patterns in IoT networks~\cite{Rezaei2019}. Moreover, DL-based methods require large volumes of high-quality labeled training data~\cite{Wang2021, Nakip2024}, which are difficult to obtain in rapidly changing IoT environments~\cite{Tong2024}, and suffer from high computational complexity and limited representation capabilities for encrypted traffic~\cite{Wei2022, Sadeghzadeh2021}. These shortcomings collectively result in low deployment efficiency in the resource-constrained environments typical of many IoT applications~\cite{Liang2022}. In addition, while some works have employed self-supervised learning methods such as masked autoencoders for traffic classification to reduce labeled data requirements~\cite{Xu2024a}, these approaches typically focus on reconstruction-based learning objectives and fail to leverage the powerful representation learning capabilities of generative models~\cite{De2022, Xie2025}, which could potentially overcome many of these limitations through more sophisticated representation learning and reduced dependence on labeled data~\cite{Xiang2023}.

\par Based on an extensive analysis of the aforementioned limitations, we seek to propose a novel framework that intelligently integrates the advantages of self-supervised learning and generative artificial intelligence (AI), thereby addressing challenges including scarcity of labeled data, difficulties in extracting meaningful features from encrypted traffic, and deployment constraints in resource-limited IoT environments. However, implementing the above framework presents several significant technical challenges. \textit{Firstly}, extracting discriminative features from network traffic visual representations requires overcoming the noisy characteristics of network data, particularly in encrypted traffic scenarios where patterns are obscured~\cite{Abbasi2021, Li2024b}. \textit{Secondly}, achieving optimal feature selection that balances classification accuracy with computational efficiency necessitates navigating a complex, high-dimensional search space that traditional optimization methods struggle to effectively explore~\cite{Song2024}. \textit{Finally}, ensuring the adaptability of the framework across diverse IoT environments with varying computational resources and traffic patterns demands solutions that can maintain performance under resource constraints while adapting to shifting data distributions~\cite{Azab2024}. Conventional approaches fail to address these interrelated challenges comprehensively, often sacrificing either performance or efficiency, and lacking the adaptability required for heterogeneous IoT deployments.

\par Accordingly, we propose DMLITE, \textit{i.e.}, diffusion model and large language model (LLM) integrated traffic embedding, which is a novel solution that directly addresses these challenges through an integrated approach combining diffusion models and LLMs. The main contributions of this paper are summarized as follows:

\begin{itemize}
    \item \textit{Generative AI-Powered Traffic Detection Architecture:} We design and implement DMLITE, a novel framework that leverages the complementary strengths of diffusion models and LLMs for IoT traffic detection. This architecture transforms raw network traffic into visual representations and employs the denoising diffusion probabilistic model (DDPM) to extract discriminative features even from encrypted traffic. To the best of our knowledge, this is the first work to integrate diffusion models and LLM-guided optimization for network traffic classification, establishing a new paradigm that moves beyond traditional discriminative approaches to network security.
  
    \item \textit{Diffusion-based Multi-level Feature Extraction:} We develop an innovative self-supervised feature extraction approach using denoising diffusion models that captures both fine-grained and abstract traffic patterns through multi-level feature fusion. Our method combines contrastive learning with representative sample selection to enable efficient fine-tuning on minimal labeled data. This approach overcomes the limitations of traditional feature extraction techniques by effectively modeling the complex distribution of network traffic data and extracting more discriminative representations, particularly valuable for encrypted traffic where subtle patterns determine classification accuracy.
  
    \item \textit{LLM-guided Adaptive Feature Selection:} We introduce a novel optimization framework that employs the DeepSeek LLM to dynamically tune particle swarm optimization (PSO) parameters for intelligent feature selection. Our dual objective function minimizes both the maximum classification error and the variance across different data distributions, ensuring robust performance across diverse deployment scenarios. This approach significantly reduces computational requirements while maintaining high classification accuracy, making the system viable for resource-constrained IoT environments where existing methods often fail to balance performance with efficiency.
  
    \item \textit{Comprehensive Performance Evaluation and Analysis:} Through extensive experiments on multiple real-world IoT traffic datasets, we demonstrate that DMLITE achieves significant improvements over the best baseline model, with an average increase of 3.7\% in classification accuracy. Moreover, ablation results further reveal important insights about the effectiveness of generative models for encrypted traffic analysis.
\end{itemize}

\par The remainder of this paper is organized as follows. Section~\ref{sec:related_work} reviews related work in network traffic detection, diffusion models, and LLMs. Section~\ref{sec:methodology} presents the detailed architecture and components of our DMLITE framework. Section~\ref{sec:results} presents and analyzes the experimental results. Finally, Section~\ref{sec:conclusion} concludes the paper.

%
\section{Related Work}
\label{sec:related_work}

\par In this section, we present a comprehensive review of existing research related to network traffic detection, focusing on the evolution from traditional methods to advanced DL and generative approaches. We categorize the literature into four main areas, which are traditional and ML-based traffic detection, DL approaches for network traffic analysis, diffusion models for representation learning, and LLMs for optimization tasks.

\subsection{Traditional and ML-based Network Traffic Detection}
\par Traditional network traffic detection methods have evolved significantly over the past decades, from simple rule-based systems to sophisticated ML approaches~\cite{Nguyen2008, Finsterbusch2014}. Initially, the port-based classification approach was used for traffic analysis, where application recognition depended on the standardized port mappings established by the Internet assigned numbers authority (IANA)~\cite{Cotton2011}. For example, the authors in~\cite{Schneider1996} proposed a TCP/IP traffic classification method based on port number correlation with applications. Similarly, the authors in~\cite{Yoon2009} introduced an application traffic classification method using fixed IP-port information automatically collected from application behavior analysis, which enables fast and accurate real-time traffic classification through simple packet header matching. Moreover, the authors in~\cite{Moore2005} and~\cite{Madhukar2006} evaluated port-based network application classification approaches. For further improving the classification accuracy of UDP traffic, the authors in~\cite{Zhang2014} investigated a component-based method, where connected half-tuples are grouped into subgraphs and classified according to the most frequently used port numbers within each group. Although port-based traffic classification methods have advantages such as faster identification speed and lower computational resources~\cite{Doroud2018}, they have become increasingly inadequate for modern network environments due to protocol camouflage and dynamic port utilization~\cite{Donato2014, Liu2021}. Specifically, some applications exploit legitimate port numbers to transmit unauthorized traffic data, such as malware concealed within HTTP streams. In addition, several applications can eschew standardized port assignments in favor of randomly selected or ephemeral ports for service delivery, as commonly observed in Voice over Internet Protocol implementations~\cite{Azab2012}. 

\par Subsequently, DPI, alternatively designated as signature-based detection methods, have been developed for traffic classification~\cite{Dainotti2012}. This type of method compares packets against predefined signature databases, thereby enabling accurate traffic classification through payload examination rather than relying on potentially misleading port associations~\cite{Azab2024}. For instance, the authors in~\cite{Sen2004} introduced a novel approach for peer-to-peer traffic identification by using application-layer signatures, which achieves accurate detection without relying on port numbers. Based on this, the authors in~\cite{Aceto2010} proposed PortLoad, a hybrid traffic classification approach that combines the efficiency and reduced privacy invasiveness of port-based methods with the accuracy of DPI techniques, achieving a better balance between classification performance and computational overhead. Further achieving faster DPI-based traffic monitoring, the authors in~\cite{AlHisnawi2016} introduced a quotient filter-based DPI approach for signature-based packet payload matching. Furthermore, the authors in~\cite{Deri2014} proposed an open-source high-speed DPI library and verified its protocol detection accuracy and efficiency in some monitoring projects. To tackle the growing challenge of encrypted communications, the authors in~\cite{Sherry2015} designed a system aimed at performing DPI operations over encrypted traffic without compromising data privacy through specialized cryptographic techniques. However, these approaches rely heavily on static rules and signatures, which makes them inadequate for detecting zero-day attacks and emerging threats in the dynamic IoT ecosystem. Moreover, they become inoperative once laws and privacy policies restrict payload access or when applications employ obfuscation and encapsulation strategies~\cite{Pacheco2019}.

\par To overcome these limitations, researchers have applied various ML techniques to network traffic classification~\cite{Nguyen2012}. The authors in~\cite{Afuwape2021}  evaluated ensemble ML classifiers, including random forest and gradient boosting, for virtual private network (VPN) and non-virtual private network (non-VPN) traffic classification, and demonstrated that ensemble methods significantly outperform single classifiers like k-nearest neighbors (KNN). Likewise, the authors in~\cite{Kumar2022} conducted a comprehensive experimental analysis of various ML algorithms for IoT network traffic classification, comparing them in terms of accuracy, speed, and training time. Recent works like~\cite{Das2022} implemented a detection framework by combining ML approaches with feature selection for network intrusion detection, and the experiment results demonstrated that carefully selected features could significantly improve detection accuracy. Moreover, the author in~\cite{Dong2021} designed a cost-sensitive multi-class support vector machine method with active learning for network traffic classification, tackling the class imbalance problem by dynamically assigning weights to applications. Additionally, the ML-based approach for encrypted traffic classification has gained traction. The authors in~\cite{Zaki2022} proposed a granular multi-label classification framework that utilizes classifier chains to classify at three levels of granular classification of encrypted network traffic for tackling the growing challenge of encrypted communications. Similarly, the authors in~\cite{Elmaghraby2024} developed three ML approaches combining neural networks and bidirectional long short-term memory (LSTM) with ensemble voting techniques for encrypted network traffic classification, achieving up to 96.8\% accuracy in classifying applications such as browsing, VoIP, file transfer, and video streaming without inspecting packet contents directly. Despite these advances, ML-based approaches are hampered by the insufficient volume of labeled network traffic samples~\cite{Shahraki2022}. Furthermore, most of them still rely heavily on domain expertise for feature engineering and struggle to adapt to the ever-changing nature of network traffic patterns~\cite{Xu2024}.

\subsection{DL Approaches for Network Traffic Analysis}

\par The limitations of traditional ML approaches have led researchers to explore DL techniques for network traffic analysis, which can automatically learn feature representations from raw data~\cite{Aceto2018, Kalwar2024}. The authors in~\cite{Wang2017} pioneered the application of convolutional neural networks (CNNs) for malware traffic classification, and this method utilized representation learning that treats raw traffic data as images, thus achieving practical accuracy requirements without requiring hand-designed features. Extending this, the authors in~\cite{Lotfollahi2020} designed Deep Packet, a DL-based framework using CNN and stacked autoencoders that integrates feature extraction and classification for both application types and traffic categories. Similarly, the authors in~\cite{Lan2022} presented a cascaded neural framework that combines a one-dimensional CNN and a bidirectional long short-term memory with self-attention to distinguish various darknet applications and protocols. For traffic analysis in IoT security, the authors in~\cite{Zhang2024} developed an automatic and efficient DL method specifically designed for resource-constrained IoT environments that maintains high detection accuracy while reducing computational overhead. Likewise, the authors in~\cite{Li2025} presented a hypergraph convolution-based framework for detecting malicious encrypted traffic in non-terrestrial IoT networks spanning satellites, unmanned aerial vehicles, and base stations. Recently, the authors in~\cite{Xiao2025} introduced RBLJAN that employs byte-label joint attention mechanisms and adversarial training to capture long-range dependencies in traffic patterns, thus enabling efficient encrypted network traffic classification. However, these approaches typically require large volumes of labeled training data, which is often scarce in rapidly evolving IoT environments, and their complex architectures demand substantial computational resources that may exceed the capabilities of many IoT devices~\cite{Xu2024a}.

\par To address the challenge of labeled data scarcity, several researchers have explored semi-supervised and self-supervised learning approaches~\cite{AbdelBasset2021, Horowicz2024}. The authors in~\cite{Dong2021a} introduced a novel semi-supervised deep reinforcement learning method that adaptively optimizes detection strategies through environmental feedback. Likewise, the authors in~\cite{Ning2022} and~\cite{Zhao2022} designed the semi-supervised learning-based ConvLaddernet and flow transformer frameworks, respectively, aimed at achieving improved classification performance when only a small quantity of annotated data is available. Recently, the authors in~\cite{Wang2024} developed a federated semi-supervised learning framework using autoencoder-based models for privacy-preserving network traffic analysis in smart home environments. In addition, the authors in~\cite{Lin2024} proposed a contrastive pre-training approach combined with semi-supervised learning to achieve robust traffic representation and mitigate classification bias. For reducing labeling dependency in traffic classification, the authors in~\cite{Zhao2023} and~\cite{Zhao2024} leveraged a masked autoencoder-based traffic transformer with multi-level flow representation to capture both local and global traffic patterns. Similarly, the authors in~\cite{Xu2024a} employed a masked autoencoder architecture and transformer-based backbone, efficiently extracting features from non-redundant traffic data. Moreover, the authors in~\cite{Xiao2024} introduced a federated self-supervised generative adversarial network-based approach that enables the recognition of traffic originating from unidentified services and produces artificial samples mimicking the distributional properties of unknown traffic patterns. In addition, the authors in~\cite{Zheng2025} developed a self-supervised learning-based multi-feature fusion framework that introduces random subset selection for data augmentation and a novel fusion mechanism to extract temporal features from traffic tables. Nevertheless, even these advanced DL approaches struggle with the dual challenges of computational efficiency and representational capacity for encrypted traffic. Specifically, most of them often fail to capture the subtle patterns that distinguish different types of encrypted traffic while maintaining the computational efficiency required for deployment in resource-constrained IoT environments.

\subsection{Diffusion Model Appliactions}

\par Diffusion models have recently emerged as powerful generative approaches that can be combined with other learning-based algorithms for control and optimization~\cite{Sun2025b, Fang2024}. For example, the authors in~\cite{Zhang2024a} explored diffusion model-based policy networks in deep reinforcement learning frameworks for energy management optimization. Similarly, the authors in~\cite{Zhao2024a} integrated discrete diffusion models with hierarchical multi-agent deep reinforcement learning for enhancing goal-conditioned navigation and sensing. Furthermore, the authors in~\cite{Zheng2025a} demonstrated how diffusion model-based approaches can enhance traditional optimization methods to achieve superior performance in complex network resource allocation and coordination problems. Likewise, the authors in~\cite{Liu2025a} explored attention-enhanced diffusion models for edge service optimization. Recent works such as~\cite{Zhang2025a} explored diffusion-enhanced reinforcement learning methods that employ the generative potential of diffusion models while integrating the powerful decision-formulation mechanisms of reinforcement learning. Additionally, the authors in~\cite{Zhang2025d} investigated the application of generative diffusion models combined with deep learning approaches for multi-objective optimization problems. Likewise, the authors in~\cite{Wang2025b} leveraged diffusion models as prediction strategies and presented a dynamic multi-objective evolutionary approach.

\par Moreover, some works have investigated applications of the diffusion model within network security~\cite{Sun2025d}. For example, the authors in~\cite{Liang2025} explored the applications of diffusion models as network optimizers and demonstrated that they can achieve flexible and efficient network performance optimization. Recent works like~\cite{Wang2025a} and~\cite{Yang2025} designed diffusion model-based approaches for Wi-Fi data processing for defending against membership inference attacks and enhancing indoor localization accuracy, respectively. Furthermore, the authors in~\cite{Zhang2025} introduced a generative diffusion model-enabled approach for secure beamforming optimization in intelligent reflecting surface-assisted IoT communications. Likewise, the authors in~\cite{Liang2025a} developed an improved twin delayed deep deterministic policy gradient algorithm based on the diffusion model to enhance unmanned aerial vehicle-enabled secure data collection in IoT networks. In addition, the authors in~\cite{Wang2025} developed a novel secure sensing system based on diffusion mode that leverages both the discrete conditional diffusion model for graph generation to optimize link activation and continuous conditional diffusion models to generate safeguarding signals, protecting users against unauthorized monitoring. To tackle data scarcity challenges in IoT malware detection, the authors in~\cite{Camerota2024} leveraged the DDPM for synthetic network traffic image generation.

\par Several studies have proposed exploiting the unique ability of diffusion models to learn complex data distributions and extract meaningful representations~\cite{Yun2024}. The authors in~\cite{Chen2024} systematically deconstructed modern diffusion models to identify their essential components for effective representation learning. Furthermore, the authors in~\cite{Xiang2023} investigated the inherent representation learning capabilities of denoising diffusion autoencoders (DDAEs) and demonstrated that these models can obtain high-quality discriminative representations within their intermediate layers through generative pre-training alone. Based on this, the authors in~\cite{Xiang2025} further proposed DDAE++, which introduces a self-conditioning mechanism that leverages the rich semantic information within diffusion networks to simultaneously improve both generative quality and discriminative performance. The authors in~\cite{Hao2024} leveraged pre-trained diffusion models for unsupervised concept extraction. Similarly, the authors in~\cite{Bandara2025} and~\cite{Sadia2024} explored the applications of diffusion models as feature extraction in sensing change detection and medical imaging, respectively. To learn complex unknown noise distributions, the authors~\cite{Zhao2024b} introduced the application of diffusion models to signal detection in near-field communication systems and presented a maximum likelihood estimation diffusion detector.

\par Despite these advances, diffusion models have seen limited application in network security contexts, with most research focusing on their generative capabilities rather than their potential for discriminative feature extraction in complex domains like encrypted network traffic. Moreover, current applications of diffusion models typically utilize only single-layer representations, failing to leverage the rich multi-level features formed during the diffusion process that could potentially capture both fine-grained and abstract patterns in network traffic data.

\begin{table*}[htb]
\renewcommand{\arraystretch}{1.5}
    \scriptsize
    \centering
    \caption{Differences between This Work and Existing Works}
    \label{tab:works comparison}
    \begin{tabular*}{\textwidth}{@{}@{\extracolsep{\fill}}cccccccccc@{}}
\toprule
\textbf{Reference}
& \begin{tabular}[c]{@{}c@{}}\textbf{ML-based} \\ \textbf{Method}\end{tabular}
& \begin{tabular}[c]{@{}c@{}}\textbf{DL-based} \\ \textbf{Method}\end{tabular}
& \begin{tabular}[c]{@{}c@{}}\textbf{Technique} \end{tabular}
& \begin{tabular}[c]{@{}c@{}}\textbf{Supervised} \\ \textbf{Learning}\end{tabular}
& \begin{tabular}[c]{@{}c@{}}\textbf{Semi-supervised} \\ \textbf{Learning}\end{tabular}
& \begin{tabular}[c]{@{}c@{}}\textbf{Self-supervised}\\ \textbf{Learning}\end{tabular}
& \begin{tabular}[c]{@{}c@{}}\textbf{Generative} \\ \textbf{Method}\end{tabular}
& \begin{tabular}[c]{@{}c@{}}\textbf{LLM}\\ \textbf{Application}\end{tabular}
& \begin{tabular}[c]{@{}c@{}}\textbf{Encrypted}\\ \textbf{Traffic}\end{tabular} \\
\hline \hline

\cite{Afuwape2021}  & $\checkmark$  & $\times$  & Ensemble classifier  & $\checkmark$   & $\times$  & $\times$ & $\times$ & $\times$ & $\checkmark$ \\ 

\cite{Dong2021}  & $\checkmark$  & $\times$  & Cost-sensitive SVM  & $\checkmark$   & $\times$  & $\times$ & $\times$ & $\times$ & $\checkmark$ \\ 

\cite{Das2022}  & $\checkmark$  & $\times$  & Ensemble Feature selection  & $\checkmark$   & $\times$  & $\times$ & $\times$ & $\times$ & $\times$ \\ 

\cite{Zaki2022}  & $\checkmark$  & $\times$  & Classifier Chain & $\checkmark$   & $\times$  & $\times$ & $\times$ & $\times$ & $\checkmark$ \\ 

\cite{Elmaghraby2024}  & $\checkmark$  & $\checkmark$  & LSTM and ensemble voting & $\checkmark$   & $\times$  & $\times$ & $\times$ & $\times$ & $\checkmark$ \\ 


\cite{Wang2017}  & $\times$ &  $\checkmark$  & CNN  & $\checkmark$   & $\times$  & $\times$ & $\times$ & $\times$ & $\times$ \\ 

\cite{Lotfollahi2020}  & $\times$ &  $\checkmark$  & CNN and stacked autoencoders  & $\checkmark$   & $\times$  & $\times$ & $\times$ & $\times$ & $\checkmark$ \\ 

\cite{Zhang2024}  & $\times$ &  $\checkmark$  & NASP  & $\checkmark$   & $\times$  & $\times$ & $\times$ & $\times$ & $\times$ \\ 

\cite{Li2025}  & $\times$ &  $\checkmark$  & Hypergraph neural networks  & $\checkmark$   & $\times$  & $\times$ & $\times$ & $\times$ & $\checkmark$ \\ 

\cite{Xiao2025}  & $\times$ &  $\checkmark$  & CNN with joint attention, GAN  & $\checkmark$   & $\times$  & $\times$ & $\checkmark$ & $\times$ & $\checkmark$ \\ 

\cite{Ning2022}  & $\times$ &  $\checkmark$  & CNN & $\times$   & $\checkmark$  & $\times$ & $\times$ & $\times$ & $\times$ \\ 

\cite{Wang2024}  & $\times$ &  $\checkmark$  & Federated learning & $\times$   & $\checkmark$  & $\times$ & $\checkmark$ & $\times$ & $\times$ \\ 

\cite{Lin2024}  & $\times$ &  $\checkmark$  & Transformer & $\times$   & $\checkmark$  & $\times$ & $\times$ & $\times$ & $\checkmark$ \\ 

\cite{Zhao2023}  & $\times$ &  $\checkmark$  & Masked autoencoder & $\times$   & $\times$  & $\checkmark$ & $\checkmark$ & $\times$ & $\times$ \\ 

\cite{Xu2024a}  & $\times$ &  $\checkmark$  & Masked autoencoder & $\times$   & $\times$  & $\checkmark$ & $\checkmark$ & $\times$ & $\times$ \\ 

\cite{Xiao2024}  & $\times$ &  $\checkmark$  & GAN and federated learning & $\times$   & $\times$  & $\checkmark$ & $\checkmark$ & $\times$ & $\times$ \\ 

\cite{Lan2022}  & $\times$ &  $\checkmark$  & CNN and LSTM  & $\checkmark$   & $\times$  & $\times$ & $\times$ & $\times$ & $\checkmark$ \\ 

\cite{Ginige2024}  & $\times$ &  $\checkmark$  & GPT-2  & $\checkmark$   & $\times$  & $\times$ & $\times$ & $\checkmark$ & $\checkmark$ \\ 

\cite{zhou2024enhancing}  & $\times$ &  $\checkmark$  & GPT-3.5-turbo  & $\checkmark$   & $\times$  & $\times$ & $\times$ & $\checkmark$ & $\times$ \\ 

\cite{chen2024merlot}  & $\times$ &  $\checkmark$  & GPT-2-base  & $\checkmark$   & $\times$  & $\times$ & $\times$ & $\checkmark$ & $\checkmark$ \\ 

\textbf{This Work}  & $\checkmark$  & $\checkmark$  & DDPM and DeepSeek  & $\times$   & $\times$  & $\checkmark$ & $\checkmark$ & $\checkmark$ & $\checkmark$ \\ 
\bottomrule
\end{tabular*}
\end{table*}

\subsection{LLM Applications in Optimization and Parameter Tuning}

\par LLMs have demonstrated remarkable capabilities across various domains, including optimization and parameter tuning tasks, due to their powerful reasoning capabilities~\cite{Li2024, OBETrans}. For example, the authors in~\cite{Zhang2024b} presented a model-centric optimization approach for democratizing LLM deployment on mobile edge networks. In addition, the authors in~\cite{Jiang2024} proposed a framework where LLMs guide the optimization process for complex engineering problems by iteratively refining solution strategies based on performance feedback. Building on this, the authors in~\cite{Du2025} introduced a reinforcement learning with LLMs interaction framework that leverages LLM-empowered generative agents to provide real-time subjective quality of experience feedback for optimizing resource allocation in distributed diffusion model services. Similarly, the authors in~\cite{Yan2025} deployed a hybrid approach based on LLM for simultaneously optimizing vehicle-to-infrastructure communication and autonomous driving policies, where LLMs optimize driving decisions through experience learning while collaborating with the double deep Q-learning algorithm for communication optimization. Furthermore, the authors in~\cite{Sun2025a} introduced an inventive paradigm exploiting LLM-enabled graphs for dynamic network optimization, demonstrating how LLMs can effectively optimize UAV trajectory planning and communication resource allocation.

\par Moreover, some works have focused on utilizing LLMs for parameter adjustment~\cite{Custode2024, Kochnev2025}. For example, the authors in~\cite{Zhang2023} and~\cite{Liu2025b} explored the use of LLMs for hyperparameter optimization in ML models, showing that language models can effectively reason about parameter relationships and suggest promising configurations based on previous performance data. Building on this, the authors in~\cite{Li2025a} utilized DeepSeek to facilitate flexible hyperparameter adjustment in deep reinforcement learning, evidencing the ability of LLMs to strengthen algorithmic outcomes through contextually-driven parameter adaptation. Similarly, the authors in~\cite{Miyake2023} proposed an LLM-enabled parameter tuning framework for robotic systems, where LLMs interpret user preferences and dynamically adjust operational parameters to optimize human-robot interaction performance in physical care tasks. Recent work like~\cite{Chen2025} introduced an LLM-based hyperparameter adaptation framework for evolutionary reinforcement learning, where the Qwen model dynamically adjusts learning rate and discount factor based on fitness. Likewise, the authors in~\cite{Mahammadli2024} proposed a novel LLM-based hyperparameter optimization framework aimed at improving parameter space exploitation. Additionally, the authors in~\cite{Li2024a} proposed a novel system aimed at employing LLMs for automatically designing dense reward functions within reinforcement learning environments, significantly improving agent learning efficiency in complex tasks through iterative reward function refinement. 

\par In evolutionary algorithms, several researchers have investigated strategies for parameter adaptation~\cite{Phan2020, Melin2013}. For instance, the authors in~\cite{Tanabe2013}, the authors designed a success-history based parameter adaptation mechanism for the differential evolution algorithm that dynamically adjusts control parameters by leveraging historical information of successful parameters. Building on this, the authors in~\cite{Viktorin2019} and~\cite{Ghosh2022} further enhanced this approach by introducing distance-based and nearest spatial neighborhood-based improvements, respectively. Likewise, the authors in~\cite{Zhou2022}  developed a parameter adaptation-enhanced ant colony optimization method with a dynamic hybrid mechanism that analyzes historical search information to adjust key parameters during optimization. Moreover, the authors in~\cite{Sun2021} employed the policy gradient method to acquire parameter control strategies from optimization experiences, which enables algorithms to autonomously adjust their behavior based on the characteristics of the problem being solved. Recent works like~\cite{Chen2024a} integrated multiple information sources to evaluate optimization progress and adjust algorithm parameters accordingly. Nevertheless, these approaches rely on predefined adaptation rules or heuristics rather than leveraging the reasoning capabilities and contextual understanding offered by LLMs, thus limiting their flexibility and generalization across diverse problem domains. Furthermore, existing optimization approaches typically focus on single-objective functions, neglecting the importance of performance robustness across different data distributions, which is a critical consideration for IoT traffic detection systems that typically operate effectively across heterogeneous deployment environments.

\par Different from these approaches, the proposed DMLITE framework aims to combine the representational power of diffusion models with the reasoning capabilities of LLMs to create a comprehensive solution for IoT traffic detection that overcomes the limitations identified in the literature. In the following, we present the details of the DMLITE framework. 

\section{DMLITE Framework: Diffusion Model and LLM Integrated Traffic Embedding}
\label{sec:methodology}

\par In this section, we introduce a novel DMLITE framework for IoT network traffic classification.  In the following, we detail the framework, beginning with an overview of the framework architecture, followed by detailed descriptions of each component, and concluding with a computational complexity analysis.

%
\subsection{Overview of DMLITE Framework}

\par As aforementioned, network traffic classification faces unique challenges, including rapidly changing communication patterns, limited labeled data for emerging threats, and resource constraints in edge computing environments~\cite{Zhang2015, Xiong2014}. Traditional classification methods often struggle with these challenges due to their reliance on manual feature engineering or their inability to adapt to new traffic patterns without extensive retraining~\cite{Zhang2013}. To address these challenges, we aim to propose an integrated approach that leverages both diffusion models and LLMs to extract and optimize discriminative features from network traffic.

\par Specifically, the designed DMLITE framework addresses these limitations through a three-stage pipeline that combines the representation power of diffusion models with the optimization capabilities of LLMs, and it consists of the following key components:

\begin{itemize}
    \item \textbf{Traffic Visual Preprocessing}: In this process, we convert raw network traffic into visual representations suitable for DL models. This transformation enables the application of powerful computer vision techniques to the domain of network traffic analysis.
    
    \item \textbf{Diffusion-based Feature Extraction}: We leverage DDPMs to extract discriminative traffic features. This approach capitalizes on the diffusion models to learn complex data distributions and capture multi-scale features.
    
    \item \textbf{LLM-guided Feature Selection Optimization}: Employing LLMs to optimize the feature selection process, reducing dimensionality while maintaining classification accuracy across diverse network environments.
\end{itemize}

\par The DMLITE framework integrates these components to enhance classification accuracy, generalization capability, and computational efficiency in IoT traffic classification tasks. In the following subsections, we detail each component and explain how they collectively address the challenges of IoT traffic classification.

%
\subsection{Traffic Visual Preprocessing}
\label{subsec:preprocessing}

\par IoT network traffic presents unique challenges for classification due to its heterogeneous nature, protocol diversity, and evolving patterns~\cite{Booij2022}. Traditional feature extraction methods often rely on domain expertise and manual feature engineering, which can be time-consuming and may not generalize well across different IoT environments~\cite{Ren2021, Kumar2022}. Visual representations offer a promising alternative, as they can preserve structural information while enabling the application of advanced DL techniques~\cite{Wang2017}.

\begin{figure*}
	\centering
	\includegraphics[width=0.9 \linewidth]{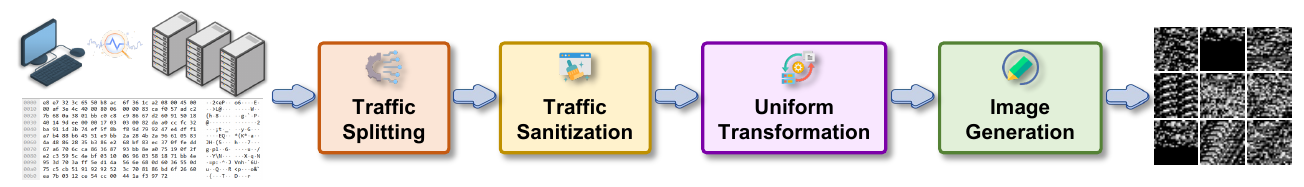}
	\caption{Sketch map of the traffic visual preprocessing.}
	\label{fig:Preprocessing}
\end{figure*}

\par As such, we transform raw network traffic into visual representations, enabling the application of powerful computer vision techniques to the traffic classification domain~\cite{Xu2024, Ding2022}. Specifically, as shown in Fig.~\ref{fig:Preprocessing}, our preprocessing module employs the USTC-TK2016 toolkit~\cite{Wang2017} to convert raw network traffic data in PCAP format into standardized image representations. In particular, the transformation process consists of four critical steps:

\begin{enumerate}
    \item \textbf{Traffic Splitting}: We divide the continuous network traffic capture into discrete traffic units based on flow granularity. Each flow $F_i$ is defined as a sequence of packets sharing the same 5-tuple (source IP, destination IP, source port, destination port, protocol), which is given by 

    \begin{equation}
        \begin{split}
            F_i = \{p_1, p_2, \ldots, p_n\} \mid \forall p_j \in F_i: \\
            \text{5-tuple}(p_j) = \text{5-tuple}(p_1).
        \end{split}
    \end{equation}

    \item \textbf{Traffic Sanitization}: To eliminate network-specific bias and ensure privacy, we perform traffic anonymization by randomizing MAC and IP addresses. Additionally, empty and duplicate samples are removed to prevent model bias, \textit{i.e.},
    \begin{equation}
        F_i' = \mathcal{A}(F_i) \mid \mathcal{A}: \text{anonymization function}.
    \end{equation}

    \item \textbf{Uniform Transformation}: Each traffic unit is trimmed or padded to a fixed length of $L = w \times h$ bytes to ensure consistency in processing. This standardization is essential for the subsequent visual transformation, which is given by
    \begin{equation}
        F_i'' = 
        \begin{cases}
            \text{trim}(F_i', L) & \text{if } |F_i'| > L, \\
            \text{pad}(F_i', L) & \text{if } |F_i'| < L, \\
            F_i' & \text{otherwise}.
        \end{cases}
    \end{equation}

    \item \textbf{Image Generation}: The byte sequence is converted into a $w \times h$ grayscale image, where each pixel corresponds to a byte value, \textit{i.e.},
    \begin{equation}
        I_i = \text{reshape}(F_i'', w, h).
    \end{equation}
\end{enumerate}

\par Note that this preprocessing approach transforms network traffic into a visual domain that preserves the inherent patterns and byte-level relationships while enabling the application of DL techniques for subsequent feature extraction and classification. The visual representation captures both local byte-level patterns and global flow-level structures, providing a rich foundation for our diffusion-based feature extraction process.

%
\subsection{Diffusion-based Feature Extraction}

\par In general, IoT traffic classification faces challenges related to feature representation, such as the presence of noise in raw traffic data, and the lack of discriminative features for distinguishing between similar traffic types. Traditional feature extraction methods often rely on handcrafted features or general-purpose neural networks that may not effectively capture the unique characteristics of network traffic~\cite{Zhang2020}. 

\par In this case, diffusion models have demonstrated remarkable capability in learning complex data distributions and extracting meaningful representations in various domains such as computer vision and natural language processing~\cite{Croitoru2023, Liu2024a}. Moreover, diffusion models have proven effective in self-supervised representation learning scenarios~\cite{Chen2024}, which makes them particularly suitable for limited labeled network traffic data in IoT environments. In particular, diffusion models demonstrate superior efficiency in modeling complex training distributions compared to other generative approaches, and have been demonstrated to extract high-quality discriminative features from their intermediate layers during the denoising process~\cite{Xiang2023}.

\begin{figure*}
    \centering
    \includegraphics[width=0.95 \linewidth]{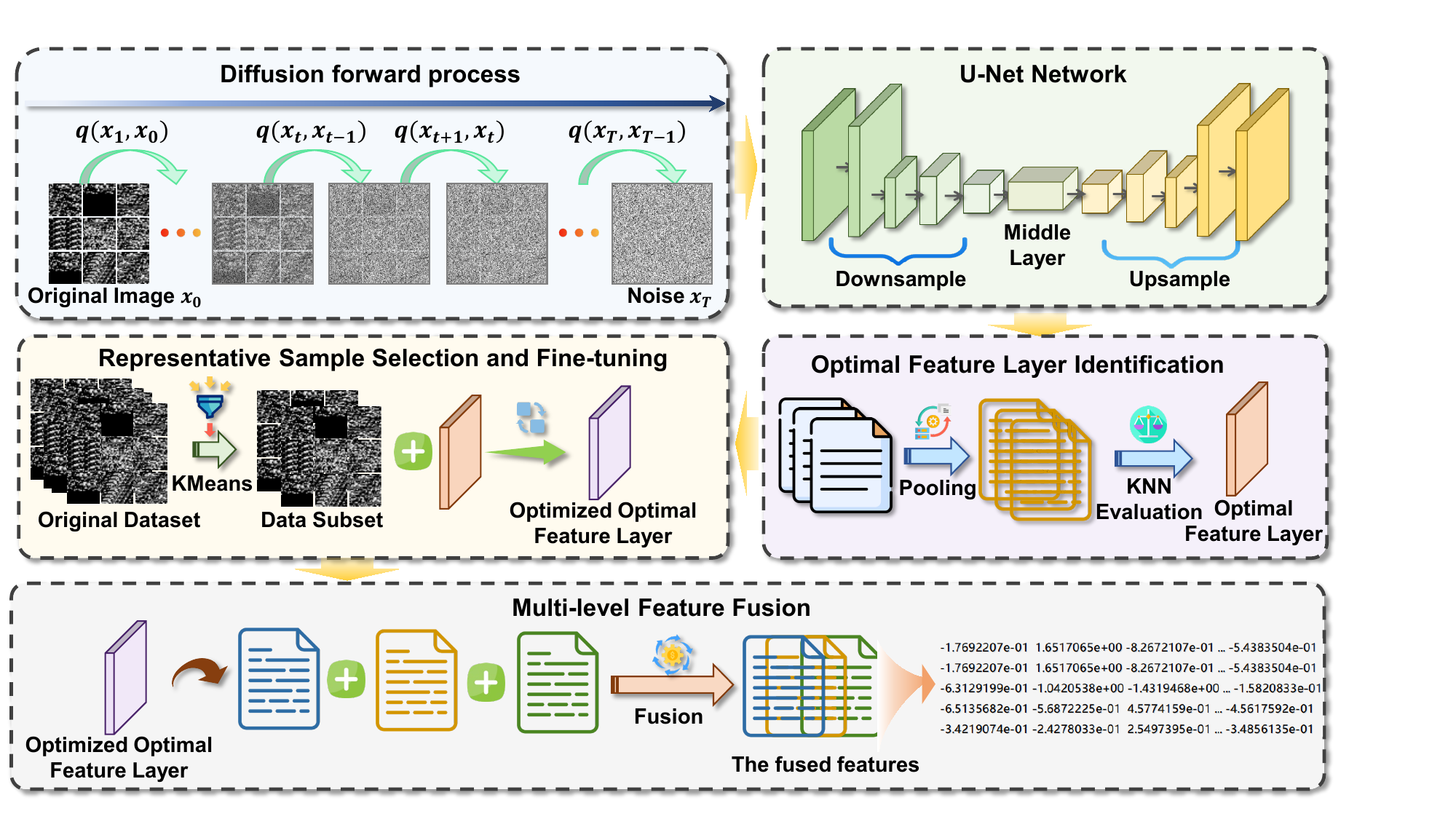}
    \caption{Comprehensive architecture of the diffusion-based multi-level feature extraction process, demonstrating: (1) the forward diffusion process that progressively adds noise to traffic images, (2) the U-Net network architecture used for denoising, (3) optimal feature layer identification through evaluation metrics, (4) representative sample selection using K-means clustering for efficient fine-tuning, and (5) multi-level feature fusion that combines features from adjacent network layers to capture both fine-grained and abstract traffic patterns.}
    \label{fig:DM_FE}
\end{figure*}
\par Thus, we introduce a novel application of DDPMs~\cite{Ho2020} for network traffic feature extraction, leveraging their ability to progressively denoise data and capture multi-scale features that are particularly valuable for distinguishing subtle differences in traffic patterns. Fig.~\ref{fig:DM_FE} illustrates the framework of our diffusion-based feature extraction approach. This approach consists of four key stages, which are DDPM model training, optimal feature layer identification, representative sample selection and fine-tuning, and multi-level feature fusion. The details are as follows.

\subsubsection{DDPM Model Training}

\par We first train a DDPM based on a U-Net architecture to learn the underlying distribution of network traffic images. The diffusion process follows a forward process that gradually adds Gaussian noise to the data and a reverse process that learns to denoise~\cite{He2025}. In particular, the forward process is given by~\cite{Ho2020}

\begin{equation}
q(x_t|x_{t-1}) = \mathcal{N}(x_t; \sqrt{1-\beta_t}x_{t-1}, \beta_t\mathbf{I}).
\end{equation}

\par Moreover, the reverse process is given by
\begin{equation}
p_\theta(x_{t-1}|x_t) = \mathcal{N}(x_{t-1}; \mu_\theta(x_t, t), \Sigma_\theta(x_t, t)),
\end{equation}

\noindent where $\beta_t$ is the noise schedule, $\mu_\theta$ and $\Sigma_\theta$ are learned by the neural network.

\par Following this, the model is trained using the variational lower bound objective as follows:
    \begin{equation}
        L(\theta) = \mathbb{E}_{x_0, \epsilon, t} \left[ \| \epsilon - \epsilon_\theta(x_t, t) \|^2 \right],
    \end{equation}
\noindent where $\epsilon$ is the random noise and $\epsilon_\theta$ is the noise predicted by the model.

\subsubsection{Optimal Feature Layer Identification}

\par Then, we evaluate different layers of the trained denoising network to identify the optimal feature extraction layer that provides the most discriminative representations. This process is important because different layers in the network capture features at varying levels of abstraction. For each layer $l$, we compute a classification performance metric $\mathcal{P}_l$ using a simple classifier as follows:
\begin{equation}
 l^* = \arg\max_l \mathcal{P}_l,
\end{equation}

\noindent where $l^*$ is the optimal layer.

\subsubsection{Representative Sample Selection and Fine-tuning}

\par In this process, we select a representative subset of the training data to fine-tune the optimal feature extraction layer by using K-means clustering~\cite{Shi2010}. This approach aims to reduce computational requirements while maintaining performance by focusing on the most informative samples, which can be given by
\begin{equation}
\mathcal{S} = \{x_i | i \in \text{KMeans}(\mathcal{F}_{l^*}(\mathcal{D}), k=n \times 0.05)\},
\end{equation}
\noindent where $\mathcal{F}_{l^*}$ represents the feature extraction at the optimal layer $l^*$, $\mathcal{D}$ is the training dataset, and $\mathcal{S}$ is the selected subset.

\par Note that the fine-tuning process employs contrastive learning loss for enhanced feature discrimination. The contrastive learning loss based on InfoNCE is defined as follows~\cite{He2020}:
\begin{equation}
\mathcal{L}_{\text{contra}} = -\frac{1}{N} \sum_{i} \log \frac{\exp(z_{i} \cdot z_{i} / \tau)}{\exp(z_{i} \cdot z_{i} / \tau) + \sum_{j \neq i} \exp(z_{i} \cdot z_{j} / \tau)},
\end{equation}

\noindent where $z_{i}$ is the normalized feature vector, $\tau$ is a temperature parameter controlling the sharpness of the distribution, and $N$ is the number of samples. This contrastive loss encourages the model to learn representations where samples from the same class are close to each other while samples from different classes are pushed apart in the feature space.

\par Following this, the optimization objective is given by
\begin{equation}
    \min_{\theta_{l^*}} \mathcal{L}_{\text{contra}}.
\end{equation}

\subsubsection{Multi-level Feature Fusion}

\par To capture features at different levels of abstraction, we concatenate features from the optimal layer with its adjacent layers (above and below) to form a comprehensive feature representation~\cite{Yang2003}. This multi-level fusion enhances the robustness of the extracted features by incorporating both fine-grained and abstract patterns, which is given by
\begin{equation}
\mathcal{F}_{\text{fused}}(x) = [\mathcal{F}_{l^*-1}(x) \oplus \mathcal{F}_{l^*}(x) \oplus \mathcal{F}_{l^*+1}(x)],
\end{equation}

\noindent where $\oplus$ denotes the concatenation operation.

\par As can be seen, this diffusion-based feature extraction approach captures the intricate patterns in network traffic while providing robust representations that are less sensitive to noise and domain shifts. By leveraging the hierarchical nature of the U-Net architecture and the denoising capabilities of diffusion models, our method extracts features that effectively distinguish between different types of IoT traffic, even when the differences are subtle or when limited labeled data is available.

%
\subsection{LLM-guided Feature Selection Optimization}
\label{subsec:selection}

\par Due to the high dimensionality of extracted features, we aim to conduct feature selection to avoid the increased computational complexity and potential overfitting. Conventional feature selection methods often rely on fixed heuristics or require extensive manual tuning of hyperparameters~\cite{Diao2015}, limiting their effectiveness across diverse IoT environments. However, the dynamic nature of IoT traffic patterns necessitates adaptive feature selection strategies that can account for varying data distributions across different network settings.

\begin{figure*}
    \centering
    \includegraphics[width=0.95\linewidth]{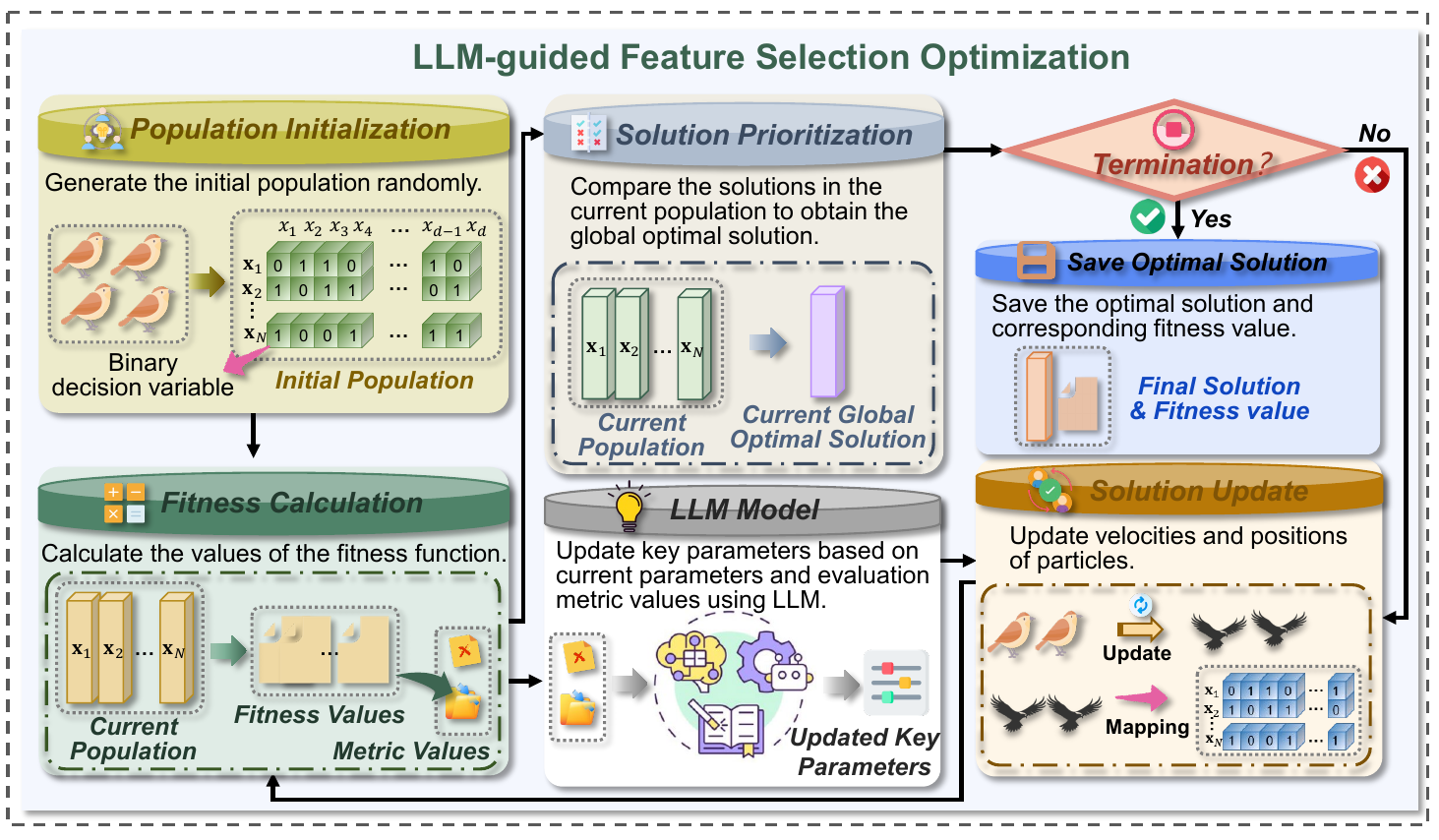}
    \caption{The LLM-guided PSO-based feature selection optimization workflow, showing how the pre-trained DeepSeek model dynamically tunes PSO parameters.}
    \label{fig:LLM_FS}
\end{figure*}

\par In this case, we integrate the LLMs into the feature selection process to develop a more intelligent and adaptive feature selection strategy. We select PSO as the foundation for feature selection due to several key advantages. Specifically, PSO demonstrates excellent convergence properties for binary optimization problems like feature selection~\cite{Zhang2014a}. Additionally, PSO offers superior computational efficiency with fewer parameters to adjust compared to genetic algorithms and other swarm intelligence algorithms~\cite{Li2025c}. Furthermore, the social learning mechanism of PSO balances global and local search through particle information exchange while maintaining population diversity throughout the optimization process, helping prevent premature convergence to suboptimal feature subsets~\cite{Abbasi2022}. By guiding the parameter optimization of the PSO algorithm, LLMs can help identify the most discriminative features while maintaining computational efficiency. Fig.~\ref{fig:LLM_FS} shows the proposed LLM-guided PSO-based feature selection algorithm. In the following, we first introduce the optimization objective during the feature selection process, then present the PSO parameter optimization, and finally detail the iterative parameter search and evaluation method.

%
%
\subsubsection{Dual Optimization Objective}

\par To ensure both high performance and robustness across different data distributions, we define a dual optimization objective that aims to minimize both the maximum classification error rate across multiple data subsets and the variance of error rates between these subsets, which is given by
\begin{equation}
\begin{aligned}
\min_{\theta} \Big\{ & \max_{i \in \{1,2,\dots,n\}} \text{Err}_i(\theta) + \\
& \lambda \cdot \text{Var}(\{\text{Err}_1(\theta), \text{Err}_2(\theta), \dots \text{Err}_n(\theta)\}) \Big\},
\end{aligned}
\label{eq:fitness}
\end{equation}

\noindent where $\text{Err}_i(\theta)$ is the classification error rate on subset $i$ with feature selection parameters $\theta$, and $\lambda$ is a weighting factor balancing the two components. This dual objective encourages the selection of features that perform well across all data subsets, rather than features that excel on some subsets but perform poorly on others.

\subsubsection{PSO Parameter Optimization}

\par Following this, we utilize the DeepSeek LLM to optimize three critical parameters of the PSO algorithm, which are inertia weight $w$, cognitive learning factor $c_1$, and social learning factor $c_2$ during optimizing the aforementioned optimization objective shown in Eq.~\eqref{eq:fitness}. In general, the standard PSO update equations are given by~\cite{Abbasi2022}

\begin{equation}
v_i^{t+1} = w \cdot v_i^t + c_1 \cdot r_1 \cdot (p_i^t - x_i^t) + c_2 \cdot r_2 \cdot (g^t - x_i^t),
\label{eq:update_v}
\end{equation}
\begin{equation}
x_i^{t+1} = x_i^t + v_i^{t+1},
\label{eq:update_x}
\end{equation}

\noindent where $v_i^t$ and $x_i^t$ are the velocity and position of particle $i$ at iteration $t$, $p_i^t$ is the best position found by particle $i$, $g^t$ is the global best position, and $r_1, r_2$ are random numbers in [0,1].

\begin{algorithm}[!htb]
\caption{LLM-Guided Parameter Optimization via DeepSeek}
\label{Algorithm:LLM-Optimization}

\KwIn{Current parameters $(w, c_1, c_2)$, recent metrics $recent\_metrics$}
\KwOut{Updated parameters $(w_{new}, c_1^{new}, c_2^{new})$}

\tcc{Step 1: Construct prompt with optimization context}
$prompt \leftarrow$ Current PSO parameters: $w$=\{$w$\}, $c_1$=\{$c_1$\}, $c_2$=\{$c_2$\}. Recent 6 iterations: \{$recent\_metrics$\} with fitness and feature counts. Objective: Minimize fitness (classification error rate + the number of selected features). Analyze convergence trend and suggest specific numerical adjustments for $w$, $c_1$, $c_2$.\;

\tcc{Step 2: Query DeepSeek for semantic guidance}
$llm\_response \leftarrow$ $llm_response$.query($prompt$) // API call to DeepSeek\;

\tcc{Step 3: Parse natural language suggestions to numerical updates}
$\Delta w, \Delta c_1, \Delta c_2 \leftarrow$ Extract parameter adjustments from $llm_response$ // Parsing uses regex patterns\;

\tcc{Step 4: Apply validated parameter updates}
$w_{new} \leftarrow w + \Delta w$\;
$c_1^{new} \leftarrow c_1 + \Delta c_1$\;
$c_2^{new} \leftarrow c_2 + \Delta c_2$\;

\tcc{Step 5: Validate parsing success and return updated parameters}
\If{Parsing failed or bounds violated}{
    $w_{new} \leftarrow w$, $c_1^{new} \leftarrow c_1$, $c_2^{new} \leftarrow c_2$ // Fallback to current values\;
}

\Return{$(w_{new}, c_1^{new}, c_2^{new})$}\;
\end{algorithm}

\par The PSO convergence analysis demonstrates that parameter configuration $(w, c_1, c_2)$ critically determines exploration-exploitation balance~\cite{Clerc2002, Trelea2003}. Different from static parameter configurations or linear parameter adjustment methods, we use the LLMs to optimize these parameters based on historical performance data and domain knowledge, which is given by 
\begin{equation}
(w^*, c_1^*, c_2^*) = \text{LLM}_{\text{optimize}}(\mathcal{H}, \mathcal{D}_{\text{val}}),
\label{eq:update_parameter}
\end{equation}
\noindent where $\mathcal{H}$ represents historical performance data and $\mathcal{D}_{\text{val}}$ is the validation dataset. Algorithm~\ref{Algorithm:LLM-Optimization} details the interaction between PSO and DeepSeek LLM. Specifically, the PSO-DeepSeek interaction is described as follows:

\par \textit{Step 1: Prompt Construction.} Designing a structured natural language query that includes current PSO parameters $(w, c_1, c_2)$, recent performance metrics, such as fitness value, classification accuracy, and feature counts over the past 6 iterations, and explicit instructions requesting specific numerical parameter adjustments to improve convergence behavior.

\par \textit{Step 2: DeepSeek API Query.} Invoking the LLM with the constructed prompt.

\par \textit{Step 3: Response Parsing.} Extracting numerical parameter changes from the natural language response of DeepSeek using regex patterns that recognize common directive formats such as increase $w$ by 0.1 or set $c_1$ to 1.8.

\par \textit{Step 4: Validated Parameter Update.} Applying the suggested changes while enforcing theoretically sound boundary constraints.

\par \textit{Step 5: Fallback Mechanism.} Reverting to previous parameter values if parsing fails or suggested values violate bounds, ensuring algorithmic robustness against ambiguous or invalid LLM responses.

\begin{algorithm}[!htb]
\caption{LLM-guided PSO Feature Selection}
\label{Algorithm:LLM-PSO}

\KwIn{Feature matrix $X$, labels $y$, population size $N$, maximum iterations $T$, data subsets $\{D_1,...,D_n\}$}
\KwOut{Selected feature subset $F_{selected}$}

\tcc{Initialize}

$w \leftarrow 0.5$, $c_1 \leftarrow 2.0$, $c_2 \leftarrow 1.5$ // Key parameters of PSO\;
$dim \leftarrow$ number of features\;
$llm\_optimizer \leftarrow$ Initialize LLM optimizer with DeepSeek API configuration\;
Randomly initialize position matrix $X$ and velocity matrix $V$\;

$X_{pb} \leftarrow X$; // Personal best positions\;
$X_{gb} \leftarrow$ zero vector; // Global best position\;
$pb_{fit} \leftarrow [\infty,...,\infty]$; // Personal best fitness\;
$gb_{fit} \leftarrow \infty$; // Global best fitness\;

$recent\_metrics \leftarrow []$; // Store recent performance metrics for LLM analysis\;

\For{$t = 1$ \KwTo $T$}{
    $X_{bin} \leftarrow \mathbb{I}(X > 0.5)$ // Map continuous positions to discrete feature selection\;
    
    \For{$i = 1$ \KwTo $N$}{ 
         Calculate particle fitness $fit_i$ across all subsets\;
        \If{$fit_i < pb_{fit}[i]$}{
            $X_{pb}[i] \leftarrow X[i]$\;
            $pb_{fit}[i] \leftarrow fit_i$\;
        }

        \If{$fit_i < gb_{fit}$}{
            $X_{gb} \leftarrow X[i]$\;
            $X_{gb\_bin} \leftarrow X_{bin}[i]$\;
            $gb_{fit} \leftarrow fit_i$\;
        }
    }
    \tcc{Record metrics for LLM-guided optimization}
    $current\_metric \leftarrow \{$iteration: $t$, best\_fitness: $gb_{fit}$, classification accuracy: $acc$,
    feature\_count: $|X_{gb\_bin}|$, params: $(w, c_1, c_2)\}$\;
    Add $current\_metric$ to $recent\_metrics$ (keep at most 6)\;
    
    \tcc{Update particle positions and velocities}
    \For{$i = 1$ \KwTo $N$}{
        Update $V_i$ and $X_i$ using Eqs.~\eqref{eq:update_v}and~\eqref{eq:update_x}\;
    }
    
    \tcc{LLM-Guided Parameter Optimization via DeepSeek}
    \If{$t > 6$}{
        Optimize PSO parameters using Algorithm~\ref{Algorithm:LLM-Optimization}\;
    }
}
\Return{Feature subset $F_{selected}$ corresponding to $X_{gb\_bin}$}\;
\end{algorithm}

\subsubsection{Iterative Parameter Search and Evaluation}
\par In what follows, the LLM iteratively refines the PSO parameters based on performance feedback, creating a feedback loop that continuously improves the feature selection process. In each iteration, the PSO algorithm with the LLM-suggested parameters selects a feature subset, which is then evaluated using the dual optimization objective, \textit{i.e.},
\begin{equation}
\mathcal{F}_{\text{selected}} = \text{PSO}_{(w^*, c_1^*, c_2^*)}(\mathcal{F}_{\text{fused}}, \mathcal{D}_{\text{train}}).
\end{equation}

\par Then, the results are fed back to the LLM, which generates improved parameter suggestions:
\begin{equation}
(w^{*'}, c_1^{*'}, c_2^{*'}) = \text{LLM}_{\text{refine}}((w^*, c_1^*, c_2^*), \mathcal{P}(\mathcal{F}_{\text{selected}})),
\label{eq:updata_LLM}
\end{equation}

\noindent where $\mathcal{P}(\mathcal{F}_{\text{selected}})$ represents the performance metrics of the selected feature subset.

\par Algorithm~\ref{Algorithm:LLM-PSO} shows the pseudocode of the LLM-guided feature selection optimization strategy. By integrating the reasoning capabilities of LLMs~\cite{Sun2025} with the exploration efficiency of PSO, this approach intelligently navigates the feature space to identify the most discriminative features for IoT traffic classification.

%
\subsection{Computational Complexity Analysis}

\par The practical deployment of IoT traffic classification systems often requires careful consideration of computational resources, especially in edge computing environments. Thus, we analyze the computational complexity of each component of the DMLITE framework and discuss optimizations for resource-constrained environments. Specifically, the computational complexity of the DMLITE framework can be analyzed for each of its major components:

\begin{enumerate}
    \item \textbf{Traffic Visual Preprocessing}: The complexity is $O(N \cdot L)$, where $N$ is the number of traffic flows and $L$ is the maximum flow length (capped at 784 bytes in our implementation). This step is relatively lightweight and can be efficiently implemented on edge devices.

    \item \textbf{Diffusion-based Feature Extraction}:
    \begin{itemize}
        \item DDPM Model Training: $O(T \cdot E \cdot N \cdot D)$, where $T$ is the number of diffusion steps, $E$ is the number of epochs, $N$ is the number of training samples, and $D$ is the dimensionality of the data. This is the most computationally intensive phase, but it is performed offline during the development stage.
        \item Optimal Layer Identification: $O(L \cdot N \cdot D)$, where $L$ is the number of layers evaluated. This is also performed offline during model development.
        \item Representative Sample Selection: $O(k \cdot N \cdot D \cdot I)$, where $k$ is the number of clusters, $N$ is the dataset size, $D$ is the feature dimension, and $I$ is the number of iterations for K-means. Our approach reduces this cost by operating on only 5\% of the training data.
        \item Fine-tuning: $O(E' \cdot N' \cdot D)$, where $E'$ is the number of fine-tuning epochs and $N'$ is the size of the representative subset (5\% of $N$). This phase is significantly more efficient than full model training due to the reduced dataset size.
    \end{itemize}

    \item \textbf{LLM-guided Feature Selection}:
    \begin{itemize}
        \item PSO Optimization: $O(P \cdot I \cdot D \cdot C)$, where $P$ is the number of particles, $I$ is the number of iterations, $D$ is the feature dimension, and $C$ is the cost of evaluating the fitness of each particle. The LLM guidance helps reduce the number of iterations required to reach an optimal solution.
        \item LLM Parameter Refinement: $O(R \cdot I_{LLM})$, where $R$ is the number of refinement rounds and $I_{LLM}$ is the inference cost of the LLM. This step is performed on a server rather than edge devices.
    \end{itemize}
\end{enumerate}

\par The overall computational complexity of the DMLITE framework is dominated by the DDPM model training phase. However, once the model is trained, the feature extraction and selection phases are relatively efficient, especially since we only use 5\% of the training data for fine-tuning. The LLM-guided optimization introduces additional computational cost, but this is offset by the improved efficiency of the resulting feature selection, leading to better classification performance with fewer features.

\begin{table*}[!htb]
\scriptsize
\centering
\caption{Dataset information processed by the USTC-TK2016 toolkit~\cite{Wang2017} in our experiments.}
\renewcommand{\arraystretch}{1.3}
\begin{tabular}{@{}ccccl@{}}
\toprule
\textbf{Dataset} & \textbf{Classes} & \textbf{Total samples} & \multicolumn{2}{l}{\textbf{Detailed class distribution}}                                                                  \\ \midrule
\multirow{4}{*}{USTC-TFC}         & \multirow{4}{*}{20}               & \multirow{4}{*}{272708}                 & \multirow{2}{*}{Malware} & Cridex: 15000 \hfill Geodo: 6000 \hfill Htbot: 13549  \hfill Miuref: 17178 \hfill Neris: 14456                              \\
                 &                  &                        &                          & Nsis-ay: 14984 \hfill Shifu: 12000 \hfill Tinba: 12653 \hfill Virut: 12982 \hfill Zeus: 15761                   \quad \quad             \\ \cmidrule(l){4-5} 
                 &                  &                        & \multirow{2}{*}{Benign}  & Bittorrent: 16335 \hfill FTP: 13804 \hfill Facetime: 12657  \hfill Gmail: 10051 \hfill MySQL: 17008            \quad \quad               \\
                 &                  &                        &                          & Outlook: 11115 \hfill SMB: 15772 \hfill Skype: 16213 \hfill Weibo: 13158 \hfill WoW: 12032                                 \\ \midrule

\multirow{2}{*}{ISCX-VPN}        & \multirow{2}{*}{12}               & \multirow{2}{*}{340339}                 & non-VPN                  & audio: 206226 \hfill chat: 11326 \hfill file: 73347 \hfill mail: 8948  \hfill streaming: 2667 \hfill voip: 4033   \quad \quad \\ \cmidrule(l){4-5} 
                 &                  &                        & VPN                      & vpn-audio: 20742 \quad vpn-chat: 8065 \quad vpn-file: 1949 \quad vpn-mail: 598 \quad vpn-streaming: 666 \quad vpn-voip: 1832  \quad \quad \\ \midrule

\multirow{6}{*}{Edge-IIoTset}         & \multirow{6}{*}{24}               & \multirow{6}{*}{570860}                 & \multirow{3}{*}{Normal traffic} & Distance: 32653 \hfill Flame\_Sensor (FS): 32687 \hfill Heart\_Rate (HR): 20842 \quad \quad \\
& & & & IR\_Receiver (IR): 32730 \hfill Modbus: 1138        \hfill phValue: 35522 \hfill Soil\_Moisture (SM): 34584 \quad \quad \\
& & & & Sound\_Sensor (SS): 32695 \hfill Temperature\_and\_Humidity (TH): 33282 \hfill Water\_Level (WL): 34330 \quad \quad \\
\cmidrule(l){4-5} 
                 &                  &                        & \multirow{3}{*}{Attack traffic}  & Backdoor: 1693 \hfill DDoS\_HTTP: 16760 \hfill DDoS\_ICMP: 60000  \hfill DDoS\_TCP: 60000 \hfill DDoS\_UDP: 60000 \quad \quad\\
                 &                  &                        &                          &  MITM:102 \hfill OS\_Fingerprinting (OS): 290 \hfill Password: 28565 \hfill  Port\_Scanning (PS): 20132 \hfill Ransomware: 1211 \quad \quad\\
                 & & & & SQL\_injection (SQL): 8790 \hfill Uploading: 15183 \hfill Vulnerability\_scanner (VS):5274 \hfill XSS: 2397 \quad \quad\\                  
                 \bottomrule
\end{tabular}
\label{tab:datasets}
\end{table*}

\par For deployment in resource-constrained IoT environments, the trained model can be further optimized through techniques such as model quantization, pruning, and knowledge distillation, reducing the overall computational requirements while maintaining classification accuracy. Additionally, the modular nature of our framework allows for flexible deployment configurations, where more computationally intensive components can be offloaded to cloud servers while lightweight inference can be performed at the edge.

\par Compared to traditional approaches that require extensive feature engineering or complex DL models that need to process raw traffic data directly, our framework offers a more balanced approach to the accuracy-efficiency trade-off, making it particularly suitable for real-world IoT security applications.

%
%
\section{Experimental Results and Analysis}
\label{sec:results}
\par In this section, we present a comprehensive evaluation of the proposed DMLITE framework. We first describe the datasets and implementation details, followed by performance comparison with state-of-the-art methods, and conclude with ablation studies to validate the contribution of different components.

\begin{table}[!htb]
\scriptsize
\renewcommand{\arraystretch}{1.3} 
\caption{Training parameters of the DMLITE implementation.}
\label{tab:parameters}
\centering
\begin{tabular}{lll}
\toprule
\textbf{Parameter} & \textbf{Description} & \textbf{Value} \\
\hline
$N_{e}$ & Training epochs & 100 \\
$\text{batch\_size}$ & Training batch size & 64 \\
$T$ & Number of diffusion timesteps & 500 \\
$d_{\text{e}}$ & Model embedding dimension & 64 \\
$t_{ex}$ & Feature extraction time step & 50 \\
\hline
\multicolumn{3}{c}{Fine-tuning stage} \\
$\rho$ & The ratio of data for fine-tuning & 0.05 \\
$N_2$ & Fine-tuning epochs & 10 \\
$\eta_2$ & Learning rate & 1e-6 \\
$c_2$ & Weight decay& 1e-2 \\
\bottomrule
\end{tabular}
\end{table}

\begin{table}[!htb]
    \centering
    \small
    \caption{Key parameter settings of baseline methods.}
    \renewcommand{\arraystretch}{1.1}
    \begin{tabular}{cll}
    \toprule
    \textbf{Method} & \textbf{Parameter} & \textbf{Value} \\
    \hline
    \multirow{5}{*}{2D-CNN~\cite{Wang2017}} & Batch size    & 64 \\
                        & Epochs        & 20 \\
                        & Learning rate & $1 \times 10^{-3}$\\
                        & Weight decay  & 0.05 \\
                        & Convolution kernels  & 5\\
    \hline
    \multirow{5}{*}{DP-CNN~\cite{Lotfollahi2020}} & Batch size    & 64 \\
                            & Epochs        & 100 \\
                            & Learning rate & $1 \times 10^{-3}$\\
                            & Dropout       & 0.05 \\
                            & Convolution kernels  & 4\\
    \hline
    \multirow{4}{*}{DP-SAE~\cite{Lotfollahi2020}} & Batch size    & 64  \\
                            & Epochs        & 100 \\
                            & Learning rate & $1 \times 10^{-3}$\\
                            & Dropout       & 0.05 \\
    \hline
    \multirow{5}{*}{RBLJAN~\cite{Xiao2025}} & Batch size    & 64   \\
                            & Epochs        & 100  \\
                            & Learning rate & $1 \times 10^{-3}$\\
                            & Dropout       & 0.5 \\
                            & Convolution kernels  & 8\\
    \hline
    \multirow{5}{*}{\makecell{MTC-MAE~\cite{Xu2024a}\\[4pt]Pre-training}} & Batch size & 64  \\
        & Epochs             & 300                \\
        & Learning rate      & $1 \times 10^{-4}$ \\
        & Weight decay       & 0.05               \\
        & The masking ratio  & 0.6                \\
    \hline
    \multirow{4}{*}{\makecell{MTC-MAE~\cite{Xu2024a}\\[4pt]Fine-tuning}}        & Batch size & 64  \\
        &Epochs              & 50                 \\
        & Learning rate      & $5 \times 10^{-4}$ \\
        & Weight decay       & 0.05                \\
    \hline
    \multirow{5}{*}{\makecell{YaTC~\cite{Zhao2023}\\[4pt]Pre-training}} & Batch size & 64  \\
        & Total step    & 150000             \\
        & Learning rate & $1 \times 10^{-3}$ \\
        & Weight decay  & 0.05               \\
        & The masking ratio  & 0.9           \\
    \hline
    \multirow{4}{*}{\makecell{YaTC~\cite{Zhao2023}\\[4pt]Fine-tuning}}        & Batch size & 64 \\
        & Epochs             & 200                \\
        & Learning rate      & $2 \times 10^{-3}$ \\
        & Weight decay       & 0.05               \\
\bottomrule
\end{tabular}
\label{tab:method_para}
\end{table}

\subsection{Dataset and Implementation}

\par To evaluate the performance of the DMLITE framework, we conduct experiments on three widely-used public network traffic datasets, \textit{i.e.}, USTC-TFC~\cite{Wang2017}, ISCX-VPN~\cite{DraperGil2016}, and Edge-IIoTset~\cite{Ferrag2022}. Table~\ref{tab:datasets} summarizes the characteristics of datasets, including the number of classes, total samples, and class distribution of samples. Specifically, the USTC-TFC dataset contains network traffic data for both malicious and benign applications. It includes 10 types of malware traffic (such as Cridex, Geodo, Htbot, Zeus, etc.) and 10 types of normal traffic activities, providing a balanced environment for evaluating traffic classification performance. Moreover, the ISCX-VPN dataset focuses on encrypted traffic classification with 12 traffic categories (6 VPN and 6 non-VPN), including audio, chat, file transfer, email service, streaming, and VoIP. This dataset is particularly valuable for evaluating performance on encrypted traffic since traditional pattern-matching techniques often fail. The Edge-IIoTset dataset comprises network traffic data collected from various edge computing and IoT devices. It includes traffic from multiple IoT device categories (such as temperature, humidity, and ultrasonic sensors) with both benign operational patterns and simulated attack traffic vectors, including DDoS, information gathering, man-in-the-middle, injection attacks, and malware. This comprehensive dataset captures the unique characteristics of resource-constrained IoT communications at network edges, which makes it particularly valuable for evaluating traffic classification algorithms in edge computing environments where processing capabilities and bandwidth are limited.

\par The proposed framework is implemented using PyTorch 2.7.1 along with CUDA 12.6 and conducted on a server equipped with NVIDIA GeForce RTX 3060 GPU. For the diffusion model component, we employ a U-Net architecture, and the diffusion process follows a cosine noise schedule, in which we use the AdamW optimizer with a learning rate of 0.001 and weight decay of 0.05. For the LLM component, we utilize a pre-trained DeepSeek model, in which the population size $N = 10$, the maximum iteration count $T = 50$, and the number of data subsets is 4. Each data subset is selected through the K-means clustering algorithm and contains approximately 5\% of the total sample. Moreover, we utilize a KNN classifier with $K=5$ as the performance evaluator for comparing feature representations extracted from different intermediate layers of the diffusion model, and the light gradient boosting machine (LightGBM) as the primary classifier for final performance evaluation and comparative experiments. Table~\ref{tab:parameters} presents the other training parameters used in the DMLITE framework.

\par As for evaluation metrics, we employ multiple complementary evaluation metrics that capture different aspects of classification performance, including accuracy (AC), precision (PR), recall (RC), and weighted F1-score (F1) as evaluation metrics. Additionally, we conduct per-class F1-score analysis to assess the effectiveness of the framework across individual traffic categories, enabling a more granular understanding of classification capabilities for specific traffic types.

\begin{table*}[!htb]
\small
\centering
\caption{Performance comparison between DMLITE and baseline algorithms on three datasets.}
\renewcommand{\arraystretch}{1.3}
\begin{tabular}{c@{\hspace{0.8em}}c@{\hspace{0.8em}}c@{\hspace{0.8em}}c@{\hspace{0.8em}}c@{\hspace{0.8em}}@{\hspace{0.8em}}c@{\hspace{0.8em}}c@{\hspace{0.8em}}c@{\hspace{0.8em}}c@{\hspace{0.8em}}@{\hspace{0.8em}}c@{\hspace{0.8em}}c@{\hspace{0.8em}}c@{\hspace{0.8em}}c}
\toprule
\multirow{2}{*}{\textbf{Model}} & \multicolumn{4}{c}{\textbf{USTC-TFC}} & \multicolumn{4}{c}{\textbf{ISCX-VPN}} & \multicolumn{4}{c}{\textbf{Edge-IIoTset}} \\
\cmidrule[\lightrulewidth](r{1.8em}){2-5} \cmidrule[\lightrulewidth](r{1.8em}){6-9} \cmidrule{10-13}
& \textbf{AC} & \textbf{PR} & \textbf{RC} & \textbf{F1} & \textbf{AC} & \textbf{PR} & \textbf{RC} & \textbf{F1} & \textbf{AC} & \textbf{PR} & \textbf{RC} & \textbf{F1} \\
\midrule
2D-CNN                  & 88.05\%          & 88.69\%          & 88.05\%          & 88.10\%          & \underline{92.43\%}    & \textbf{92.90\%} & \underline{92.43\%}    & \underline{92.57\%}    & 79.95\%          & 82.76\%          & 79.95\%          & 80.86\%          \\
DP-CNN                  & 68.22\%          & 72.52\%          & 68.22\%          & 65.02\%          & 84.29\%          & 84.59\%          & 84.29\%          & 83.23\%          & 39.15\%          & 44.17\%          & 39.15\%          & 34.41\%          \\
DP-SAE                  & 37.33\%          & 41.68\%          & 37.33\%          & 34.62\%          & 53.17\%          & 62.24\%          & 53.17\%          & 51.05\%          & 11.68\%          & 25.89\%          & 11.68\%          & 10.66\%          \\
RBLJAN                  & 86.74\%          & 87.60\%          & 86.74\%          & 85.73\%          & 89.63\%          & 89.83\%          & 89.63\%          & 89.49\%          & 50.71\%          & 65.21\%          & 50.71\%          & 45.60\%          \\
\midrule
YaTC                    & \underline{96.33\%}    & \underline{96.44\%}    & \underline{96.33\%}    & \underline{96.33\%}    & 87.68\%          & 87.53\%          & 87.68\%          & 87.60\%          & \underline{97.06\%}    & \underline{97.99\%}    & \underline{97.06\%}    & \underline{96.91\%}    \\
MTC-MAE                 & 92.10\%          & 94.77\%          & 92.10\%          & 89.43\%          & 66.67\%          & 77.78\%          & 66.67\%          & 53.34\%          & 89.97\%          & 92.22\%          & 89.97\%          & 86.01\%          \\
\midrule
\textbf{DMLITE} & \textbf{98.87\%} & \textbf{98.88\%} & \textbf{98.87\%} & \textbf{98.87\%} & \textbf{92.61\%} & \underline{92.65\%}    & \textbf{92.61\%} & \textbf{92.63\%} & \textbf{99.83\%} & \textbf{99.83\%} & \textbf{99.83\%} & \textbf{99.83\%} \\
\bottomrule
\end{tabular}
\label{tab:comparsion_results}
\end{table*}

\begin{table}[!htb]
    \centering
    \small
    \caption{Training time of different models (s).}
    \renewcommand{\arraystretch}{1.3}
    \begin{tabular}{cccc}
    \toprule
    \textbf{Model} & \textbf{USTC-TFC} & \textbf{ISCX-VPN} & \textbf{Edge-IIoTset}\\
    \hline
        2D-CNN  & 2611.75  & 2131.06  & 4059.90  \\
        DP-CNN  & 4674.83  & 4251.73  & 8261.12  \\
        DP-SAE  & 2324.84  & 2022.44  & 4463.43  \\
        RBLJAN  & 37809.12 & 24804.05 & 43403.27 \\
        YaTC    & 11039.82 & 10622.12 & 34145.02 \\
        MTC-MAE & 22797.50 & 17957.72 & 43968.08 \\
        \hline
        DMLITE  & 4339.35  & 5418.86  & 28660.60 \\ 
    \bottomrule
    \end{tabular}
    \label{tab:train_time}
\end{table}

\subsection{Comparison with Baselines}

\par To verify the performance of the proposed DMLITE framework, we conduct comprehensive experiments comparing it with several state-of-the-art baselines, including supervised learning methods (\textit{i.e.}, 2D-CNN~\cite{Wang2017}, Deep Packet~\cite{Lotfollahi2020}, and RBLJAN~\cite{Xiao2025}), and self-supervised learning approaches (\textit{i.e.}, YaTC~\cite{Zhao2023} and MTC-MAE~\cite{Xu2024a}). Note that DP-CNN and DP-SAE are Deep Packet with the one-dimensional CNN and the stacked autoencoder, respectively, and the key parameter settings of these baselines are shown in Table~\ref{tab:method_para}.

\begin{table*}[!htb]
\scriptsize
\caption{Category-wise F1 comparison between DMLITE and baseline algorithms on the USTC-TFC dataset.}
\centering
\renewcommand{\arraystretch}{1.3}
\begin{tabular}{ccccccccccc}
\toprule
\textbf{Model}
 & \textbf{Cridex} & \textbf{Geodo} & \textbf{Htbot} & \textbf{Miuref} & \textbf{Neris} & \textbf{Nsis-ay} & \textbf{Shifu} & \textbf{Tinba} & \textbf{Virut} & \textbf{Zeus} \\
\hline
2D-CNN  & 85.63\%  & 81.63\% & 79.27\% & 81.30\% & 90.01\% & 88.00\%  & 87.55\% & 92.77\%  & 89.95\% & 96.43\%  \\
DP-CNN  & 59.35\%  & 52.65\% & 57.44\% & 15.14\% & 71.19\% & 66.51\%  & 27.72\% & 19.68\%  & 69.29\% & 42.91\%  \\
DP-SAE  & 44.42\%  & 37.05\% & 49.20\% & 4.85\%  & 16.84\% & 40.82\%  & 8.94\%  & 4.91\%   & 41.46\% & 16.42\%  \\
RBLJAN  & 69.32\%  & 27.53\% & 78.47\% & 41.71\% & 92.12\% & 99.47\%  & 92.74\% & 60.56\%  & 89.00\% & 96.71\%  \\
YaTC    & \textbf{100.00\%} & 99.39\% & 91.67\% & 98.11\% & 73.44\% & \textbf{100.00\%} & 99.36\% & \textbf{100.00\%} & 77.78\% & \textbf{100.00\%} \\
MTC-MAE & 99.97\%  & 99.58\% & 98.58\% & 98.74\% & 71.00\% & 96.05\%  & 98.16\% & 99.35\%  & 0.00\%  & 99.74\%  \\
\midrule
DMLITE  & 99.97\%  & \textbf{99.96\%} & \textbf{99.80\%} & \textbf{99.85\%} & \textbf{91.54\%} & 99.37\%  & \textbf{99.84\%} & 99.94\%  & \textbf{90.22\%} & 99.83\% \\
\bottomrule
\end{tabular}\\[0.08cm]
\begin{tabular}{ccccccccccc|c}
\toprule
\textbf{Model} & \textbf{Bittorrent} & \textbf{FTP} & \textbf{Facetime} & \textbf{Gmail} & \textbf{MySQL} & \textbf{Outlook} & \textbf{SMB} & \textbf{Skype} & \textbf{Weibo} & \textbf{WoW} & \textbf{VAE} \\
\hline
2D-CNN  & 88.01\%  & 93.93\%  & 88.23\%  & 89.00\%  & 85.17\%  & 89.39\%  & 88.28\%  & 88.74\%  & 88.90\% & 90.59\%  & 88.01\%  \\
DP-CNN  & 42.47\%  & 80.01\%  & 98.35\%  & 73.14\%  & 56.80\%  & 9.93\%   & 91.47\%  & 66.56\%  & \textbf{99.93\%} & 93.15\%  & 57.92\%  \\
DP-SAE  & 61.17\%  & 72.44\%  & 44.60\%  & 21.88\%  & 12.18\%  & 24.76\%  & 30.64\%  & 6.20\%   & 47.90\% & 39.85\%  & 30.88\%  \\
RBLJAN  & \textbf{100.00\%} & 99.99\%  & 99.98\%  & 99.99\%  & \textbf{100.00\%} & 99.99\%  & 99.98\%  & 99.95\%  & 99.98\% & 99.97\% & 86.71\%  \\
YaTC    & \textbf{100.00\%} & \textbf{100.00\%} & \textbf{100.00\%} & \textbf{100.00\%} & \textbf{100.00\%} & \textbf{100.00\%} & \textbf{100.00\%} & \textbf{100.00\%} & 99.37\% & \textbf{100.00\%} & 96.80\%  \\
MTC-MAE & 99.28\%  & 0.00\%   & 99.85\%  & 87.80\%  & 99.38\%  & 98.90\%  & 93.40\%  & 99.19\%  & 99.59\% & 99.48\%  & 86.24\%  \\
\midrule
DMLITE  & \textbf{100.00\%} & \textbf{100.00\%} & 99.96\%  & 99.91\%  & \textbf{100.00\%} & 99.90\%  & \textbf{100.00\%} & \textbf{100.00\%} & 99.96\% & \textbf{100.00\%}  & \textbf{98.95\%}  \\
\bottomrule
\end{tabular}
\label{tab:ustc}
\end{table*}

\begin{table*}[!htb]
\scriptsize
\caption{Category-wise F1 comparison between DMLITE and baseline algorithms on the ISCX-VPN dataset.}
\centering
\renewcommand{\arraystretch}{1.3} 
\begin{tabular}{c@{\hspace{1em}}c@{\hspace{1em}}c@{\hspace{1em}}c@{\hspace{1em}}c@{\hspace{1em}}c@{\hspace{1em}}c@{\hspace{1em}}c@{\hspace{1em}}c@{\hspace{1em}}c@{\hspace{1em}}c@{\hspace{1em}}c@{\hspace{1em}}c|@{\hspace{1em}}c}
\toprule
\textbf{Model} & \textbf{Audio}  & \textbf{Chat}   & \textbf{File}   & \textbf{Mail}   & \textbf{ST} & \textbf{Voip}   & \textbf{Vpn-audio} & \textbf{Vpn-chat} & \textbf{Vpn-file} & \textbf{Vpn-mail} & \textbf{Vpn-ST} & \textbf{Vpn-voip} & \textbf{VAE}\\
\hline
2D-CNN  & 95.18\% & 78.45\% & 89.01\% & 71.42\% & 80.30\% & 84.11\% & 99.08\%  & 96.17\% & 85.48\% & 75.00\% & 86.57\% & 94.35\% & 86.26\%  \\
DP-CNN  & 92.96\% & 83.59\% & 93.79\% & 54.10\% & 91.13\% & 46.10\% & 95.83\%  & 86.78\% & 90.06\% & 75.79\% & 98.20\% & 68.96\% & 81.44\%  \\
DP-SAE  & 48.53\% & 44.51\% & 47.47\% & 10.74\% & 62.70\% & 18.48\% & 78.77\%  & 61.44\% & 47.04\% & 54.61\% & 98.95\% & 20.01\%  & 49.44\%  \\
RBLJAN  & \textbf{98.02\%} & 88.39\% & \textbf{99.22\%} & 69.47\% & \textbf{99.68\%} & 64.20\% & 98.62\%  & 94.64\% & 95.02\% & 96.21\% & \textbf{99.84\%} & 64.61\%  & 88.99\%  \\
YaTC    & 89.48\% & 89.00\% & 70.95\% & 90.48\% & 93.49\% & 94.07\% & \textbf{100.00\%} & \textbf{99.32\%} & \textbf{96.92\%} & \textbf{98.33\%} & 97.78\% & 99.12\%  & 93.24\%  \\
MTC-MAE & 80.17\% & 0.00\%  & 0.00\%  & 0.00\%  & 0.00\%  & 0.00\%  & 78.33\%  & 0.00\%  & 0.00\%  & 0.00\%  & 0.00\%  & 0.00\%   & 13.21\%  \\
\midrule
DMLITE  & 94.37\% & \textbf{93.37\%} & 84.57\% & \textbf{92.81\%} & 94.07\% & \textbf{95.04\%} & 99.93\%  & 97.89\% & 92.54\% & 94.83\% & 93.02\% & \textbf{99.73\%}  & \textbf{94.35\%}  \\
\bottomrule
\end{tabular}
\label{tab:iscx}
\end{table*}

\par Table~\ref{tab:comparsion_results} demonstrates that DMLITE achieves superior performance across the three datasets. The reason may be that the diffusion-based multi-level feature extraction mechanism effectively captures meaningful features through the U-Net architecture during the progressive denoising process. Moreover, compared to supervised methods, the enhanced performance exhibited by DMLITE and other self-supervised approaches, \textit{i.e.}, YaTC and MTC-MAE, validates the efficacy of self-supervised learning paradigms in network traffic classification tasks. Meanwhile, the self-supervised learning paradigm of DMLITE effectively overcomes the critical challenge of labeled data in rapidly evolving IoT environments. Additionally, the well-balanced precision and recall scores obtained by DMLITE underscore its effectiveness in reducing both false positives and false negatives, which is crucial for practical traffic classification systems that focus on classification stability. However, the relatively lower accuracy on the ISCX-VPN dataset reveals potential challenges when handling certain types of complex encrypted VPN traffic patterns. The reason may be that VPN encryption introduces additional layers of obfuscation that make traffic patterns more homogeneous, thereby reducing the discriminative power of extracted features.

\par The computational overhead comparison is illustrated in Table~\ref{tab:train_time}, showing training duration requirements for all three datasets. Although 2D-CNN exhibits superior computational efficiency, its classification accuracy proves insufficient when handling intricate network traffic patterns, as evidenced by our performance assessments. DMLITE strikes an optimal balance between computational demands and classification quality, requiring reasonable training durations while substantially outperforming self-supervised alternatives such as YaTC and MTC-MAE in terms of efficiency.

\par The detailed category-specific performance analysis is illustrated in Tables~\ref{tab:ustc},~\ref{tab:iscx}, and \ref{tab:edge}, which provide an exhaustive evaluation comparing DMLITE against baseline approaches across three distinct datasets. The experimental results indicate that baseline methods exhibit significant performance fluctuations, failing entirely to identify certain traffic categories, particularly when dealing with encrypted network traffic or complex network patterns. In contrast, the proposed DMLITE framework achieves remarkably stable and superior performance across all traffic categories throughout the three experimental datasets. Notably, self-supervised approaches such as MTC-MAE exhibit substantially diminished detection capabilities for multiple traffic categories where DMLITE demonstrates exceptional proficiency, thereby validating the superior feature extraction capabilities. Furthermore, these experimental results show the superiority of DMLITE in preserving classification effectiveness across heterogeneous network environments, especially for security-critical deployment scenarios where uniform cross-category performance reliability is paramount.

\subsection{Ablation Study}
\par To evaluate the effectiveness of each component in the proposed DMLITE framework, we conduct an ablation study by incrementally adding key components to the baseline model. Specifically, we compare and analyze four variants of our approach, which are as follows:
\begin{itemize}
    \item \textbf{DMLITE-1}: Utilizes only the optimal feature extraction layer from the diffusion model without any enhancements.
    \item \textbf{DMLITE-2}: Extends DMLITE-1 by incorporating the contrastive learning-based fine-tuning strategy.
    \item \textbf{DMLITE-3}: Builds upon DMLITE-2 by adding the multi-layer feature fusion mechanism.
    \item \textbf{DMLITE}: Our complete framework that enhances DMLITE-3 with the LLM-guided feature selection optimization.
\end{itemize}

\par
\begin{figure*}[!htbp]
\begin{minipage}[t]{1\linewidth}
  \centering
	\includegraphics[width=.33\linewidth]{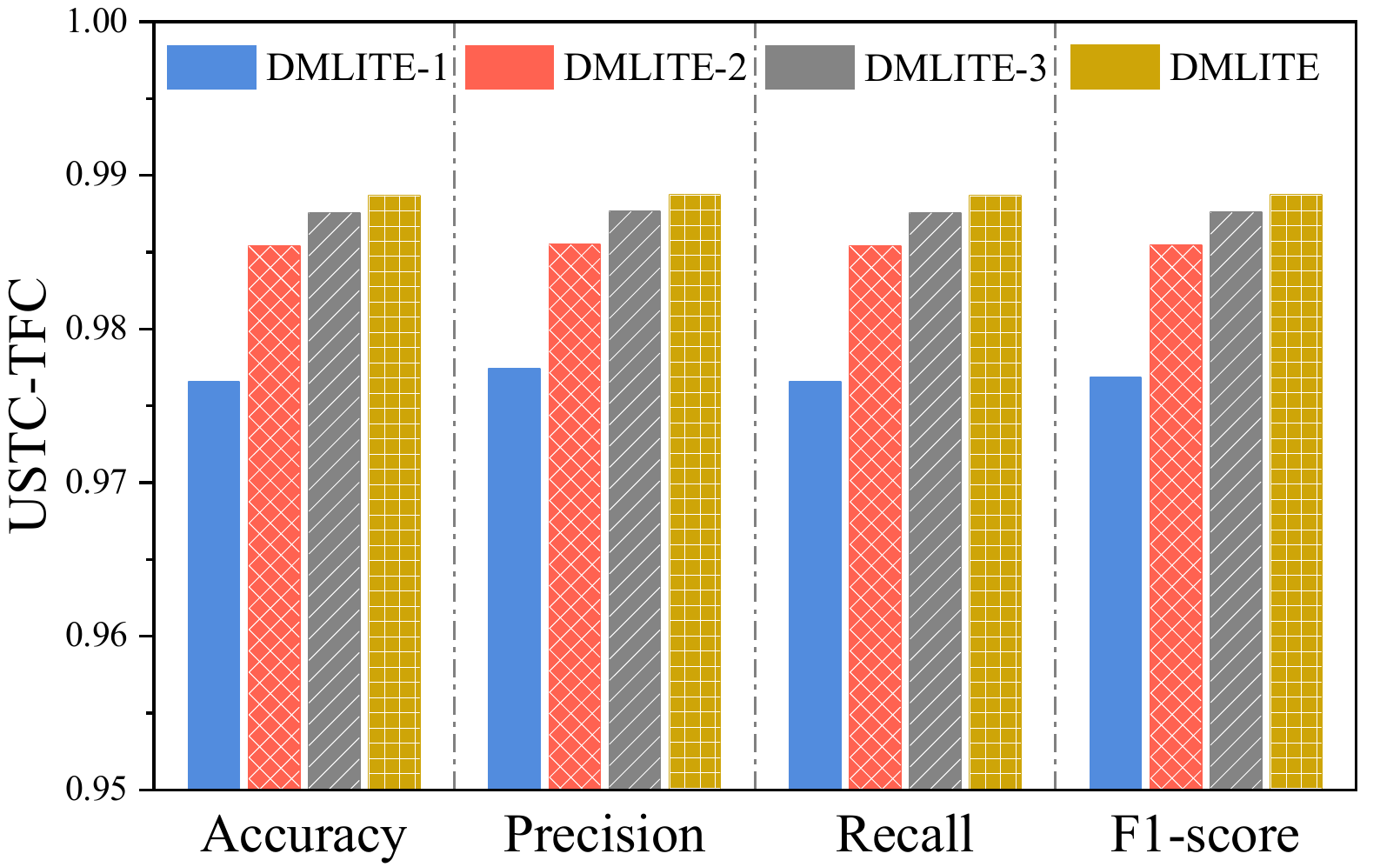}\hfill
    \includegraphics[width=.33\linewidth]{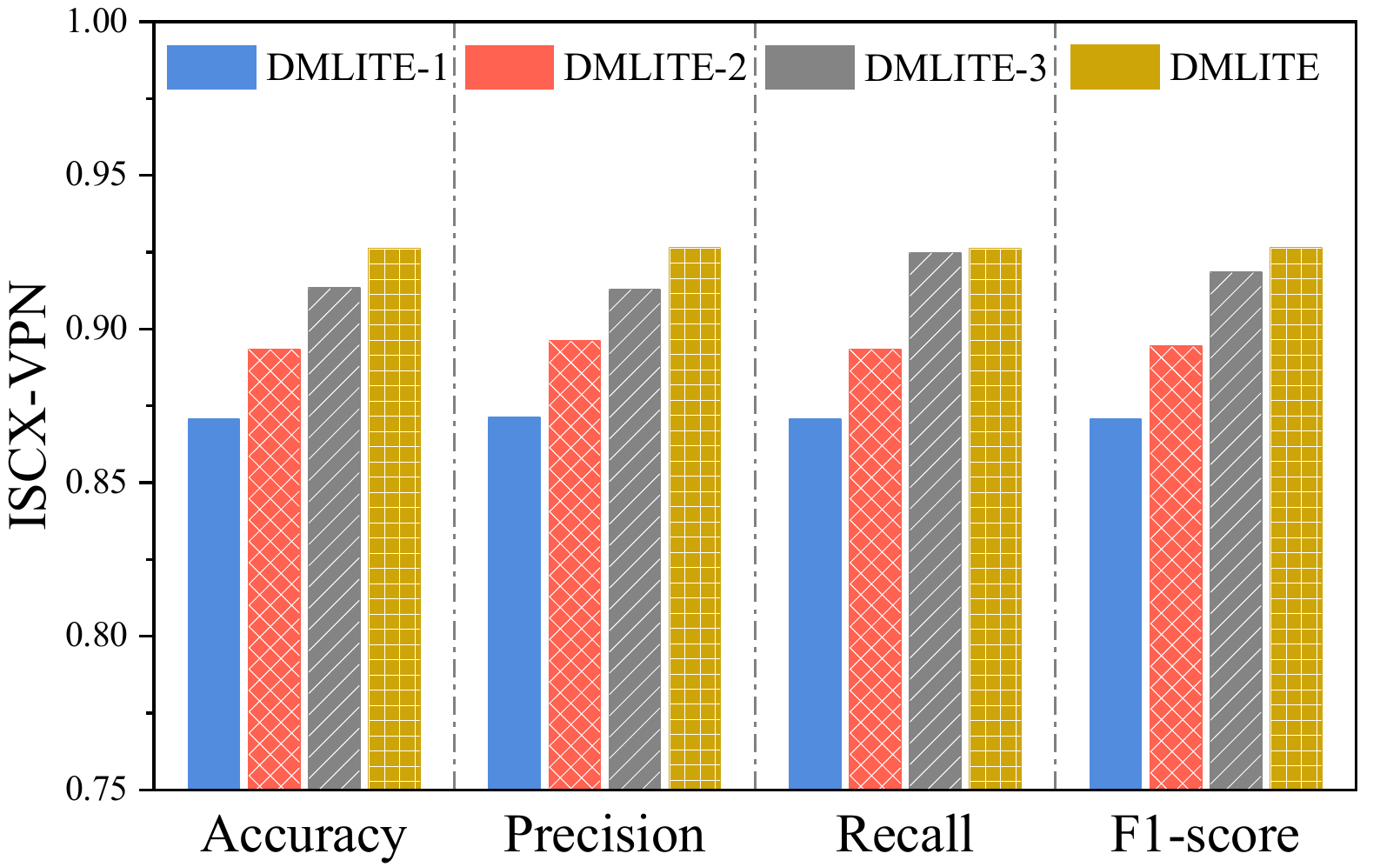}\hfill
	\includegraphics[width=.33\linewidth]{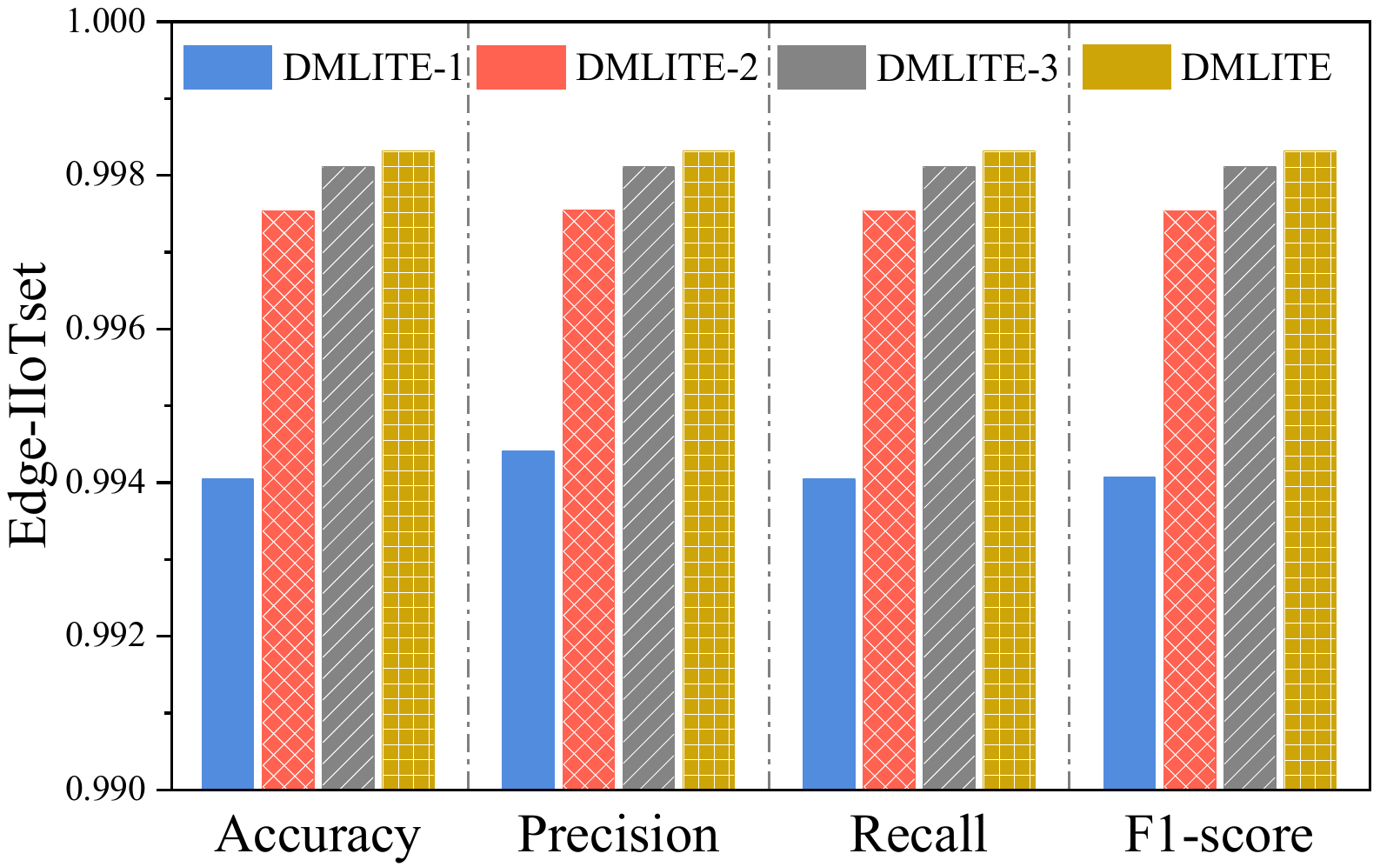}\hfill
    \caption{Performance comparison on three datasets, including accuracy, precision, recall, and F1-score.}
    \label{fig:ablation_results}
\end{minipage}
\end{figure*}

 Fig.~\ref{fig:ablation_results} presents the comparative performance of these variants on both the USTC-TFC, ISCX-VPN, and Edge-IIoTset datasets across four evaluation metrics, including accuracy, precision, recall, and F1-score. Although these components of the DMLITE framework may introduce the complexity of deployment and maintenance, experimental results show that each component delivers performance improvements across three datasets, which validates the necessity and effectiveness of this multi-stage design approach.

\par The variant DMLITE-1 shows remarkable classification performance on both USTC-TFC and Edge-IIoTset datasets, which validates that diffusion models possess capabilities for extracting meaningful representations from complex network traffic patterns. The reason can be explained by the diffusion model's unique denoising methodology, which progressively learns to reconstruct traffic data from noise, thereby acquiring a deep understanding of the underlying data distribution characteristics that effectively distinguish between different traffic categories.

\par The incorporation of contrastive learning-based fine-tuning mechanisms produces the most significant performance boost among all proposed enhancements. This improvement is particularly pronounced on the challenging ISCX-VPN dataset, where we document a substantial 2.27\% accuracy improvement compared to the baseline DMLITE-1 configuration. This enhancement can be attributed to the introduced contrastive learning framework, which learn discriminative embeddings by maximizing similarity between positive pairs while minimizing similarity between negative samples, thus refining the feature space organization for improved classification boundaries.

\par Finally, LLM-guided feature selection optimization is integrated into the model as a final enhancement. Although this component introduces additional computational overhead, it achieves consistent performance refinements across all evaluation metrics and experimental datasets. This component leverages the reasoning capabilities of LLMs to identify and prioritize the most informative features, resulting in a more robust classification framework. The comprehensive ablation analysis demonstrates that our proposed multi-component architecture achieves optimal performance through the synergistic combination of diffusion-based representation learning, contrastive fine-tuning, and intelligent feature optimization strategies.

\begin{table*}[!htb]
\scriptsize
\caption{Category-wise F1 comparison between DMLITE and baseline algorithms on the Edge-IIoTset dataset.}
\centering
\renewcommand{\arraystretch}{1.3}
\begin{tabular}{c@{\hspace{0.8em}}c@{\hspace{0.8em}}c@{\hspace{0.8em}}c@{\hspace{0.8em}}c@{\hspace{0.8em}}c@{\hspace{0.8em}}c@{\hspace{0.8em}}c@{\hspace{0.8em}}c@{\hspace{0.8em}}c@{\hspace{0.8em}}c@{\hspace{0.8em}}c@{\hspace{0.8em}}c}

\toprule
\textbf{Model} & \textbf{Distance} & \textbf{FS} & \textbf{HR} & \textbf{IR} & \textbf{Modbus} & \textbf{phValue} & \textbf{SM} & \textbf{SS} & \textbf{TH} & \textbf{WL} & \textbf{Backdoor} & \textbf{DDoS-HTTP}\\
\hline
2D-CNN  & 77.20\% & 81.58\%  & 75.77\% & 86.06\% & 26.72\% & 83.63\% & 72.06\% & 83.71\% & 82.93\% & 76.95\% & 29.09\% & 77.40\% \\
DP-CNN  & 3.26\%  & 5.54\%   & 4.01\%  & 2.07\%  & 30.02\% & 13.84\% & 7.46\%  & 3.40\%  & 66.95\% & 3.43\%  & 35.67\% & 98.74\% \\
DP-SAE  & 3.35\%  & 3.22\%   & 2.76\%  & 1.86\%  & 22.90\% & 3.04\%  & 3.04\%  & 3.19\%  & 38.95\% & 3.16\%  & 8.85\%  & 12.92\% \\
RBLJAN  & 25.87\% & 30.68\%  & 5.44\%  & 5.06\%  & 58.32\% & 10.22\% & 7.30\%  & 15.09\% & 69.34\% & 12.47\% & \textbf{99.72\%} & 99.49\% \\
YaTC    & 68.43\% & \textbf{100.00\%} & \textbf{99.95\%} & 99.95\% & \textbf{92.17\%} & 99.87\% & \textbf{99.97\%} & \textbf{99.98\%} & 80.61\% & \textbf{99.96\%} & 97.06\% & \textbf{99.85\%} \\
MTC-MAE & 99.74\% & 96.90\%  & 96.24\% & 99.03\% & 0.00\%  & 79.10\% & 93.82\% & 99.17\% & 95.35\% & 94.62\% & 0.00\%  & 0.00\%  \\
\midrule
DMLITE  & \textbf{99.95\%} & 99.94\%  & 99.93\% & \textbf{99.97\%} & 89.57\% & \textbf{99.92\%} & 99.83\% & 99.91\% & \textbf{99.56\%} & 99.77\% & 98.19\% & 99.79\% \\
\bottomrule
\end{tabular}\\[0.08cm]
\begin{tabular}{c@{\hspace{0.8em}}c@{\hspace{0.8em}}c@{\hspace{0.8em}}c@{\hspace{0.8em}}c@{\hspace{0.8em}}c@{\hspace{0.8em}}c@{\hspace{0.8em}}c@{\hspace{0.8em}}c@{\hspace{0.8em}}c@{\hspace{0.8em}}c@{\hspace{0.8em}}c@{\hspace{0.8em}}c|@{\hspace{0.8em}}c}
\toprule
\textbf{Model} & \textbf{DDoS-ICMP} & \textbf{DDoS-TCP} & \textbf{DDoS-UDP} & \textbf{MITM} & \textbf{OS} & \textbf{Password} & \textbf{Port} & \textbf{Ransomware} & \textbf{SQL} & \textbf{Uploading} & \textbf{VS} & \textbf{XSS} & \textbf{VAE}\\
\hline
2D-CNN  & 82.57\% & 83.36\%  & 92.61\%  & 10.42\% & 19.66\% & 83.07\% & 78.20\% & 26.79\% & 79.63\% & 61.20\% & 59.85\% & 64.61\%  & 66.46\%  \\
DP-CNN  & 0.00\%  & 0.00\%   & 0.00\%   & 0.00\%  & 12.87\% & 84.28\% & 5.68\%  & 96.93\% & 37.81\% & 85.78\% & 97.30\% & 27.74\% & 30.12\%  \\
DP-SAE  & 0.00\%  & 0.00\%   & 0.00\%   & 2.60\%  & 4.48\%  & 9.43\%  & 1.14\%  & 16.25\% & 16.86\% & 7.54\%  & 26.30\% & 4.80\%  & 8.19\%  \\
RBLJAN  & 0.00\%  & 30.81\%  & 0.00\%   & 64.13\% & 96.51\% & 98.48\% & 55.87\% & 93.08\% & 99.39\% & 98.73\% & \textbf{99.93\%} & 96.85\% & 53.03\%  \\
YaTC    & 99.96\% & \textbf{100.00\%} & \textbf{100.00\%} & 34.62\% & \textbf{98.25\%} & \textbf{99.83\%} & \textbf{99.80\%} & \textbf{97.07\%} & 99.66\% & 99.84\% & 98.77\% & \textbf{99.16\%} & 94.37\%  \\
MTC-MAE & 92.73\% & 99.04\%  & 99.72\%  & 0.00\%  & 0.00\%  & 80.30\% & 95.78\% & 0.00\%  & 0.00\%  & 0.00\%  & 0.00\%  & 0.00\%  & 55.06\%  \\
\midrule
DMLITE  & \textbf{99.98\%} & 99.97\%  & 99.98\%  & \textbf{70.59\%} & 96.67\% & 99.72\% & 99.63\% & 96.67\% & \textbf{99.83\%} & \textbf{99.93\%} & 98.96\% & \textbf{99.16\%} & \textbf{97.81\%} \\
\bottomrule
\end{tabular}
\label{tab:edge}
\end{table*}

\begin{figure}
    \centering
    \includegraphics[width=0.65\linewidth]{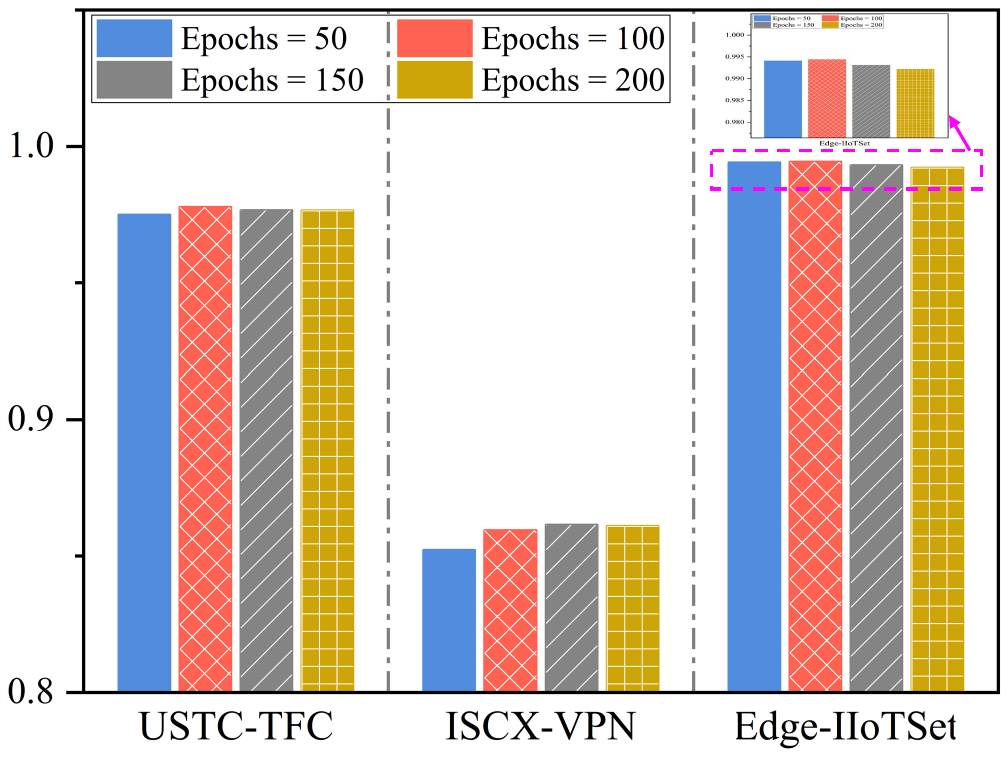}
    \caption{Classification accuracy achieved across the three datasets using different training epochs.}
    \label{fig:epochs}
\end{figure}

\subsection{Feature Extraction Analysis}
\par The diffusion-based feature extraction process represents a critical component of the DMLITE framework. Thus, we conduct experiments to investigate how varying training epochs affects the quality of extracted features in our diffusion-based feature extraction module. Specifically, we evaluate four different training configurations with epochs set to 50, 100, 150, and 200, respectively, while maintaining other hyperparameters constant. For each configuration, we extract features from the optimal layer and evaluate their classification accuracy across all three benchmark datasets.

\par As shown in Fig.~\ref{fig:epochs}, classification accuracy improves as training epochs increase from 50 to 100, with accuracy gains of 0.30\%, 0.86\%, and 0.03\% on USTC-TFC, ISCX-VPN, and Edge-IIoTset, respectively. This improvement can be attributed to the progressive learning of the underlying traffic data distribution by the diffusion model, where additional training iterations allow the denoising network to better capture fine-grained features through the U-Net architecture. However, further increasing epochs from 100 to 150 shows mixed results. Specifically, while ISCX-VPN achieves a marginal improvement of 0.23\%, both USTC-TFC and Edge-IIoTset experience slight performance degradation of 0.12\% and 0.13\%, respectively. When extending training to 200 epochs, we observe consistent performance degradation of approximately 0.01\%-0.09\% across all datasets, which suggests potential overfitting where the model begins to memorize training-specific patterns rather than generalizing to discriminative traffic features. Although ISCX-VPN achieves its peak performance at 150 epochs, the marginal 0.23\% improvement comes at the cost of nearly 50\% additional training time. Thus, we select 100 epochs as the default configuration in the DMLITE framework, as it provides an optimal balance between feature extraction quality and computational efficiency across diverse datasets.

\par Moreover, the extraction timestep parameter ($t_{ex}$) directly influences the quality and characteristics of the extracted features. This parameter determines at which point in the diffusion process features are captured, thereby affecting the balance between low-level details and high-level semantic information. Specifically, following the diffusion formulation, the noised version of $x_0$ at timestep $t$ is defined as $x_t = \alpha_t x_0 + \sigma_t \epsilon$, where $\alpha_t$ and $\sigma_t$ control how much original semantic information is preserved versus how much stochastic noise is introduced~\cite{Xiang2023}. In this case, different timesteps activate distinct levels of semantic information within the denoising network. Moreover, a more optimal timestep creates an implicit information bottleneck that compresses high-level semantics into compact, linearly-separable features, with alignment and uniformity metrics demonstrating that features extracted at optimal timesteps achieve balanced semantic consistency and discriminative specificity~\cite{Xiang2025}. Thus, to systematically evaluate the impact of this parameter on classification performance, we conducted experiments with three different extraction timestep values, which are $t_{ex} = 10$, $t_{ex} = 25$, and $t_{ex} = 50$.

\begin{figure}
    \centering
    \includegraphics[width=0.55\linewidth]{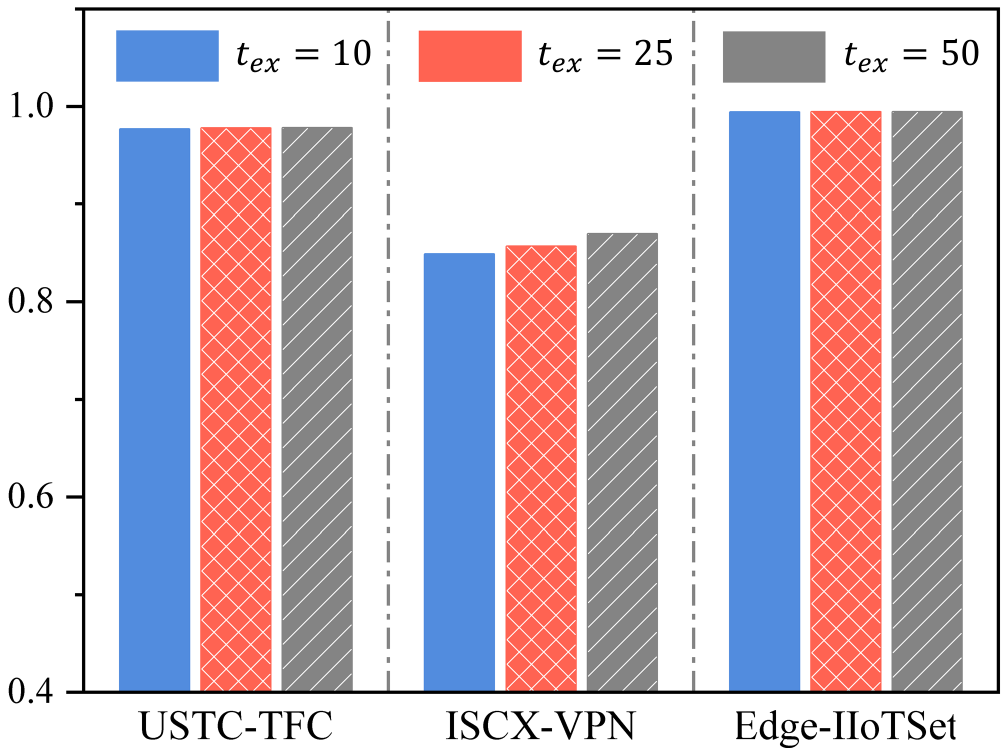}
    \caption{Classification accuracy achieved across the three datasets using different extraction timesteps.}
    \label{fig:FE_comparison}
\end{figure}

\par Fig.~\ref{fig:FE_comparison} presents the classification accuracy achieved across the three datasets using different extraction timesteps. The experimental results show that increasing the extraction timestep generally leads to improved classification performance across all datasets. In particular, this significant enhancement on the more challenging encrypted traffic dataset suggests that higher extraction timesteps enable the diffusion model to capture more abstract and discriminative features that are particularly valuable for distinguishing between complex encrypted traffic patterns.

\par These findings indicate that features extracted at later stages of the diffusion process contain richer semantic information beneficial for classification tasks, especially for complex traffic patterns. The progressive improvement with increasing extraction timesteps can be attributed to the diffusion model's ability to gradually refine its understanding of the underlying data distribution through iterative denoising steps. However, it is worth noting that the performance gains begin to plateau, particularly for the USTC-TFC and Edge-IIoTset datasets, suggesting an optimal range for the extraction timestep parameter rather than an indefinite improvement with increasing values.

\section{Conclusion}
\label{sec:conclusion}
\par This paper proposes the DMLITE framework that combines diffusion models with DeepSeek to tackle network traffic classification challenges in computationally restricted IoT environments. The proposed DMLITE first introduces a self-supervised diffusion-based hierarchical feature extraction method that identifies multiscale traffic characteristics within encrypted communication channels without extensive labeled datasets. Then, an LLM-guided adaptive feature selection method that dynamically optimizes the feature space while preserving computational efficiency through intelligent parameter tuning. Extensive experimental validation across diverse benchmark datasets confirms DMLITE delivers substantial performance enhancements compared to baselines, and achieves exceptional classification accuracy across different traffic categories while dramatically reducing computational overhead. However, the framework presents higher computational complexity during the DDPM training phase and introduces deployment complexity due to its multi-component architecture, requiring careful coordination between different modules.

\par Future research directions include investigating lightweight diffusion architectures tailored for resource-constrained environments through neural architecture search or knowledge distillation techniques to further reduce computational overhead while maintaining feature quality. Additionally, developing automatic timestep selection mechanisms based on dataset characteristics would enhance the generalizability of the method across diverse traffic distributions without manual hyperparameter tuning. Finally, extending the framework to handle continual learning scenarios where new traffic patterns emerge dynamically would address the evolving nature of IoT network environments and enhance long-term deployment viability in real-world applications.

\bibliographystyle{IEEEtran}
\bibliography{TC}

@Article{Nguyen2008,
  author  = {Thuy T. T. Nguyen and Grenville J. Armitage},
  journal = {{IEEE} Commun. Surv. Tutorials},
  title   = {A survey of techniques for internet traffic classification using machine learning},
  year    = {2008},
  number  = {1-4},
  pages   = {56--76},
  volume  = {10},
}

@Article{Finsterbusch2014,
  author  = {Michael Finsterbusch and Chris Richter and Eduardo Rocha and Jean{-}Alexander M{\"{u}}ller and Klaus Hanssgen},
  journal = {{IEEE} Commun. Surv. Tutorials},
  title   = {A Survey of Payload-Based Traffic Classification Approaches},
  year    = {2014},
  number  = {2},
  pages   = {1135--1156},
  volume  = {16},
}

@Article{Pacheco2019,
  author  = {Fannia Pacheco and Ernesto Exposito and Mathieu Gineste and C{\'{e}}dric Baudoin and Jos{\'{e}} Aguilar},
  journal = {{IEEE} Commun. Surv. Tutorials},
  title   = {Towards the Deployment of Machine Learning Solutions in Network Traffic Classification: A Systematic Survey},
  year    = {2019},
  number  = {2},
  pages   = {1988--2014},
  volume  = {21},
}

@Article{Cotton2011,
  author  = {Michelle Cotton and Lars Eggert and Joe Touch and Magnus Westerlund and Stuart Cheshire},
  journal = {{RFC}},
  title   = {Internet Assigned Numbers Authority {(IANA)} Procedures for the Management of the Service Name and Transport Protocol Port Number Registry},
  year    = {2011},
  pages   = {1--33},
  volume  = {6335},
}

@InProceedings{Yoon2009,
  author    = {Sung{-}Ho Yoon and Jin{-}Wan Park and Jun{-}Sang Park and Young{-}Seok Oh and Myung{-}Sup Kim},
  booktitle = {Proc. {APNOMS}},
  title     = {Internet Application Traffic Classification Using Fixed {IP}-Port},
  year      = {2009},
  pages     = {21--30},
  volume    = {5787},
}

@Article{Schneider1996,
  author  = {Schneider, Patrick},
  journal = {Division Of Applied Sciences, Cambridge, MA},
  title   = {{TCP/IP} traffic classification based on port numbers},
  year    = {1996},
  number  = {5},
  pages   = {1--6},
  volume  = {2138},
}

@InProceedings{Moore2005,
  author    = {Andrew W. Moore and Konstantina Papagiannaki},
  booktitle = {Proc. {PAM}},
  title     = {Toward the Accurate Identification of Network Applications},
  year      = {2005},
  pages     = {41--54},
  volume    = {3431},
}

@InProceedings{Madhukar2006,
  author    = {Alok Madhukar and Carey L. Williamson},
  booktitle = {Proc. {MASCOTS}},
  title     = {A Longitudinal Study of {P2P} Traffic Classification},
  year      = {2006},
  pages     = {179--188},
}

@InProceedings{Azab2012,
  author    = {Azab, Ahmad and Watters, Paul and Layton, Robert},
  booktitle = {Proc. {IEEE CTC}},
  title     = {Characterising network traffic for skype forensics},
  year      = {2012},
  pages     = {19--27},
}

@InProceedings{Sen2004,
  author    = {Subhabrata Sen and Oliver Spatscheck and Dongmei Wang},
  booktitle = {Proc. {WWW}},
  title     = {Accurate, scalable in-network identification of {P2P} traffic using application signatures},
  year      = {2004},
  pages     = {512--521},
}

@InProceedings{Aceto2010,
  author    = {Aceto, Giuseppe and Dainotti, Alberto and De Donato, Walter and Pescap{\'e}, Antonio},
  booktitle = {Proc. {IEEE INFOCOM}},
  title     = {{PortLoad}: Taking the best of two worlds in traffic classification},
  year      = {2010},
  pages     = {1--5},
}

@InProceedings{Zhang2014,
  author    = {Qianli Zhang and Yunlong Ma and Jilong Wang and Xing Li},
  booktitle = {Proc. {IEEE APNOMS}},
  title     = {{UDP} traffic classification using most distinguished port},
  year      = {2014},
  pages     = {1--4},
}

@Article{AlHisnawi2016,
  author  = {Mohammad Al{-}Hisnawi and Mahmood Ahmadi},
  journal = {{IEEE} Commun. Lett.},
  title   = {Deep Packet Inspection Using Quotient Filter},
  year    = {2016},
  number  = {11},
  pages   = {2217--2220},
  volume  = {20},
}

@InProceedings{Deri2014,
  author    = {Luca Deri and Maurizio Martinelli and Tomasz Bujlow and Alfredo Cardigliano},
  booktitle = {Proc. {IWCMC}},
  title     = {{nDPI}: Open-source high-speed deep packet inspection},
  year      = {2014},
  pages     = {617--622},
}

@Article{Sherry2015,
  author  = {Justine Sherry and Chang Lan and Raluca Ada Popa and Sylvia Ratnasamy},
  journal = {{IACR} Cryptol. ePrint Arch.},
  title   = {{BlindBox}: Deep Packet Inspection over Encrypted Traffic},
  year    = {2015},
  pages   = {264},
}

@Article{Azab2024,
  author  = {Ahmad Azab and Mahmoud Khasawneh and Saed Alrabaee and Kim{-}Kwang Raymond Choo and Maysa Sarsour},
  journal = {Digit. Commun. Networks},
  title   = {Network traffic classification: Techniques, datasets, and challenges},
  year    = {2024},
  number  = {3},
  pages   = {676--692},
  volume  = {10},
}

@Article{Kumar2022,
  author  = {Rakesh Kumar and Mayank Swarnkar and Gaurav Singal and Neeraj Kumar},
  journal = {{IEEE} Internet Things J.},
  title   = {{IoT} Network Traffic Classification Using Machine Learning Algorithms: An Experimental Analysis},
  year    = {2022},
  number  = {2},
  pages   = {989--1008},
  volume  = {9},
}

@Article{Zaki2022,
  author  = {Faiz Zaki and Firdaus Afifi and Shukor Abd Razak and Abdullah Gani and Nor Badrul Anuar},
  journal = {Comput. Networks},
  title   = {{GRAIN:} Granular multi-label encrypted traffic classification using classifier chain},
  year    = {2022},
  pages   = {109084},
  volume  = {213},
}

@Article{Das2022,
  author  = {Saikat Das and Sajal Saha and Annita Tahsin Priyoti and Etee Kawna Roy and Frederick T. Sheldon and Anwar Haque and Sajjan G. Shiva},
  journal = {{IEEE} Trans. Netw. Serv. Manag.},
  title   = {Network Intrusion Detection and Comparative Analysis Using Ensemble Machine Learning and Feature Selection},
  year    = {2022},
  number  = {4},
  pages   = {4821--4833},
  volume  = {19},
}

@Article{Afuwape2021,
  author  = {Afeez Ajani Afuwape and Ying Xu and Joseph Henry Anajemba and Gautam Srivastava},
  journal = {Comput. Stand. Interfaces},
  title   = {Performance evaluation of secured network traffic classification using a machine learning approach},
  year    = {2021},
  pages   = {103545},
  volume  = {78},
}

@Article{Elmaghraby2024,
  author  = {Elmaghraby, Reham T and Aziem, Nada M Abdel and Sobh, Mohammed A and Bahaa-Eldin, Ayman M},
  journal = {Ain Shams Engineering Journal},
  title   = {Encrypted network traffic classification based on machine learning},
  year    = {2024},
  number  = {2},
  pages   = {102361},
  volume  = {15},
}

@Article{Lan2022,
  author  = {Jinghong Lan and Xudong Liu and Bo Li and Yanan Li and Tongtong Geng},
  journal = {Comput. Secur.},
  title   = {{DarknetSec}: A novel self-attentive deep learning method for darknet traffic classification and application identification},
  year    = {2022},
  pages   = {102663},
  volume  = {116},
}

@Article{Dong2021,
  author  = {Shi Dong},
  journal = {Expert Syst. Appl.},
  title   = {Multi class {SVM} algorithm with active learning for network traffic classification},
  year    = {2021},
  pages   = {114885},
  volume  = {176},
}

@Article{Shahraki2022,
  author  = {Amin Shahraki and Mahmoud Abbasi and Amir Taherkordi and Anca Delia Jurcut},
  journal = {{IEEE} Trans. Cogn. Commun. Netw.},
  title   = {Active Learning for Network Traffic Classification: A Technical Study},
  year    = {2022},
  number  = {1},
  pages   = {422--439},
  volume  = {8},
}

@Article{Kalwar2024,
  author        = {Jawad Hussain Kalwar and Sania Bhatti},
  journal       = {CoRR},
  title         = {Deep Learning Approaches for Network Traffic Classification in the Internet of Things ({IoT}): A Survey},
  year          = {2024},
  volume        = {abs/2402.00920},
  archiveprefix = {arXiv},
}

@Article{Lotfollahi2020,
  author  = {Mohammad Lotfollahi and Mahdi Jafari Siavoshani and Ramin Shirali Hossein Zade and Mohammdsadegh Saberian},
  journal = {Soft Comput.},
  title   = {Deep packet: A novel approach for encrypted traffic classification using deep learning},
  year    = {2020},
  number  = {3},
  pages   = {1999--2012},
  volume  = {24},
}

@InProceedings{Wang2017,
  author    = {Wei Wang and Ming Zhu and Xuewen Zeng and Xiaozhou Ye and Yiqiang Sheng},
  booktitle = {Proc. {ICOIN}},
  title     = {Malware traffic classification using convolutional neural network for representation learning},
  year      = {2017},
  pages     = {712--717},
}

@Article{Xiao2025,
  author  = {Xi Xiao and Shuo Wang and Guangwu Hu and Qing Li and Kelong Mao and Xiapu Luo and Bin Zhang and Shutao Xia},
  journal = {{IEEE} Trans. Dependable Secur. Comput.},
  title   = {{RBLJAN:} Robust Byte-Label Joint Attention Network for Network Traffic Classification},
  year    = {2025},
  number  = {3},
  pages   = {2161--2178},
  volume  = {22},
}

@Article{Zhang2024,
  author  = {Xixi Zhang and Liang Hao and Guan Gui and Yu Wang and Bamidele Adebisi and Hikmet Sari},
  journal = {{IEEE} Internet Things J.},
  title   = {An Automatic and Efficient Malware Traffic Classification Method for Secure Internet of Things},
  year    = {2024},
  number  = {5},
  pages   = {8448--8458},
  volume  = {11},
}

@Article{Li2025,
  author    = {Li, Xuzeng and Zhang, Tao and Wang, Jian and Han, Zhen and Wang, Nan and Fan, Shaohua and Du, Hongyang and Kang, Jiawen and Liu, Jiqiang and Niyato, Dusit},
  journal   = {{IEEE} Internet Things J.},
  title     = {Protect {NTN-IoT} Security by Malicious Traffic Detection: A Multi-Dimensional Hypergraph Learning Approach},
  year      = {2025},
  publisher = {IEEE},
}

@Article{Xu2024a,
  author  = {Ke Xu and Xixi Zhang and Yu Wang and Tomoaki Ohtsuki and Bamidele Adebisi and Hikmet Sari and Guan Gui},
  journal = {{IEEE} Internet Things J.},
  title   = {Self-Supervised Learning Malware Traffic Classification Based on Masked Autoencoder},
  year    = {2024},
  number  = {10},
  pages   = {17330--17340},
  volume  = {11},
}

@Article{Ning2022,
  author  = {Jinhui Ning and Guan Gui and Yu Wang and Jie Yang and Bamidele Adebisi and Song Ci and Haris Gacanin and Fumiyuki Adachi},
  journal = {{IEEE} Internet Things J.},
  title   = {Malware Traffic Classification Using Domain Adaptation and Ladder Network for Secure Industrial Internet of Things},
  year    = {2022},
  number  = {18},
  pages   = {17058--17069},
  volume  = {9},
}

@InProceedings{Zhao2023,
  author    = {Ruijie Zhao and Mingwei Zhan and Xianwen Deng and Yanhao Wang and Yijun Wang and Guan Gui and Zhi Xue},
  booktitle = {Proc. {AAAI}},
  title     = {Yet Another Traffic Classifier: A Masked Autoencoder Based Traffic Transformer with Multi-Level Flow Representation},
  year      = {2023},
  pages     = {5420--5427},
}

@Article{Wang2024,
  author  = {Zixuan Wang and Zeyi Li and Mengyi Fu and Yingchun Ye and Pan Wang},
  journal = {J. Syst. Archit.},
  title   = {Network traffic classification based on federated semi-supervised learning},
  year    = {2024},
  pages   = {103091},
  volume  = {149},
}

@InProceedings{Zhao2022,
  author    = {Ruijie Zhao and Xianwen Deng and Zhicong Yan and Jun Ma and Zhi Xue and Yijun Wang},
  booktitle = {Proc. {ACM KDD}},
  title     = {{MT-FlowFormer}: A Semi-Supervised Flow Transformer for Encrypted Traffic Classification},
  year      = {2022},
  pages     = {2576--2584},
}

@Article{Lin2024,
  author  = {Xinjie Lin and Longtao He and Gaopeng Gou and Jing Yu and Zhong Guan and Xiang Li and Juncheng Guo and Gang Xiong},
  journal = {Comput. Secur.},
  title   = {{CETP:} A novel semi-supervised framework based on contrastive pre-training for imbalanced encrypted traffic classification},
  year    = {2024},
  pages   = {103892},
  volume  = {143},
}

@Article{AbdelBasset2021,
  author  = {Mohamed Abdel{-}Basset and Hossam Hawash and Ripon K. Chakrabortty and Michael J. Ryan},
  journal = {{IEEE} Internet Things J.},
  title   = {Semi-Supervised Spatiotemporal Deep Learning for Intrusions Detection in {IoT} Networks},
  year    = {2021},
  number  = {15},
  pages   = {12251--12265},
  volume  = {8},
}

@Article{Zhao2024,
  author  = {Ruijie Zhao and Mingwei Zhan and Xianwen Deng and Fangqi Li and Yanhao Wang and Yijun Wang and Guan Gui and Zhi Xue},
  journal = {{IEEE/ACM} Trans. Netw.},
  title   = {A Novel Self-Supervised Framework Based on Masked Autoencoder for Traffic Classification},
  year    = {2024},
  number  = {3},
  pages   = {2012--2025},
  volume  = {32},
}

@Article{Xiao2024,
  author  = {Yong Xiao and Rong Xia and Yingyu Li and Guangming Shi and Diep N. Nguyen and Dinh Thai Hoang and Dusit Niyato and Marwan Krunz},
  journal = {{IEEE} Trans. Mob. Comput.},
  title   = {Distributed Traffic Synthesis and Classification in Edge Networks: A Federated Self-Supervised Learning Approach},
  year    = {2024},
  number  = {2},
  pages   = {1815--1829},
  volume  = {23},
}

@Article{Zheng2025,
  author  = {Xuan Zheng and Xiuli Ma and Lifu Xu and Yanliang Jin and Chun Ke},
  journal = {{IEEE} Trans. Netw. Serv. Manag.},
  title   = {Augmentation and Fusion: Multi-Feature Fusion-Based Self-Supervised Learning Approach for Traffic Tables},
  year    = {2025},
  number  = {3},
  pages   = {2647--2662},
  volume  = {22},
}

@Article{Horowicz2024,
  author  = {Eyal Horowicz and Tal Shapira and Yuval Shavitt},
  journal = {{IEEE} Trans. Netw. Serv. Manag.},
  title   = {Self-Supervised Traffic Classification: Flow Embedding and Few-Shot Solutions},
  year    = {2024},
  number  = {3},
  pages   = {3054--3067},
  volume  = {21},
}

@InProceedings{Bandara2025,
  author    = {Wele Gedara Chaminda Bandara and Nithin Gopalakrishnan Nair and Vishal M. Patel},
  booktitle = {Proc. {IEEE/CVF WACV}},
  title     = {{DDPM-CD:} Denoising Diffusion Probabilistic Models as Feature Extractors for Remote Sensing Change Detection},
  year      = {2025},
  pages     = {5250--5262},
}

@Article{Yun2024,
  author        = {Kwan Yun and Youngseo Kim and Kwanggyoon Seo and Chang Wook Seo and Junyong Noh},
  journal       = {CoRR},
  title         = {Representative Feature Extraction During Diffusion Process for Sketch Extraction with One Example},
  year          = {2024},
  volume        = {abs/2401.04362},
  archiveprefix = {arXiv},
}

@Article{Wang2025,
  author  = {Jiacheng Wang and Hongyang Du and Yinqiu Liu and Geng Sun and Dusit Niyato and Shiwen Mao and Dong In Kim and Xuemin Shen},
  journal = {{IEEE} Trans. Inf. Forensics Secur.},
  title   = {Generative {AI} Based Secure Wireless Sensing for {ISAC} Networks},
  year    = {2025},
  pages   = {5195--5210},
  volume  = {20},
}

@Article{Liang2025,
  author  = {Ruihuai Liang and Bo Yang and Pengyu Chen and Xianjin Li and Yifan Xue and Zhiwen Yu and Xuelin Cao and Yan Zhang and M{\'{e}}rouane Debbah and H. Vincent Poor and Chau Yuen},
  journal = {{IEEE} Internet Things J.},
  title   = {Diffusion Models as Network Optimizers: Explorations and Analysis},
  year    = {2025},
  number  = {10},
  pages   = {13183--13193},
  volume  = {12},
}

@Article{Liang2025a,
  author  = {Shuang Liang and Minhao Yin and Wenwen Xie and Zemin Sun and Jiahui Li and Jiacheng Wang and Hongyang Du},
  journal = {{IEEE} Internet Things J.},
  title   = {{UAV}-Enabled Secure Data Collection and Energy Transfer in {IoT} via Diffusion-Model-Enhanced Deep Reinforcement Learning},
  year    = {2025},
  number  = {10},
  pages   = {13455--13468},
  volume  = {12},
}

@Article{Zhang2025,
  author  = {Jing Zhang and Zheng Liu and Xin Feng and Hongwei Yang and Shuang Liang},
  journal = {{IEEE} Internet Things J.},
  title   = {Enhanced Secure Beamforming for {IRS}-Assisted {IoT} Communication Using a Generative-Diffusion-Model-Enabled Optimization Approach},
  year    = {2025},
  number  = {10},
  pages   = {13398--13414},
  volume  = {12},
}

@InProceedings{Wang2025a,
  author    = {Ningning Wang and Tianya Zhao and Shiwen Mao and Xuyu Wang},
  booktitle = {Proc. {IEEE INFOCOM}},
  title     = {Privacy-Preserving {Wi-Fi} Data Generation via Differential Privacy in Diffusion Models},
  year      = {2025},
  pages     = {1--10},
}

@InProceedings{Camerota2024,
  author    = {Chiara Camerota and Lorenzo Pappone and Tommaso Pecorella and Flavio Esposito},
  booktitle = {Proc. {IEEE CNSM}},
  title     = {Addressing Data Security in {IoT}: {Minimum} Sample Size and Denoising Diffusion Models for Improved Malware Detection},
  year      = {2024},
  pages     = {1--7},
}

@Article{Liu2025a,
  author  = {Yaju Liu and Siyuan Li and Xi Lin and Xiuzhen Chen and Gaolei Li and Yuchen Liu and Bolin Liao and Jianhua Li},
  journal = {{IEEE} Trans. Cogn. Commun. Netw.},
  title   = {{QoS}-Aware Multi-{AIGC} Service Orchestration at Edges: {An} Attention-Diffusion-Aided {DRL} Method},
  year    = {2025},
  number  = {2},
  pages   = {1078--1090},
  volume  = {11},
}

@Article{Zhang2025a,
  author  = {Zheng Zhang and Jingjing Wang and Jianrui Chen and Hang Fu and Ziheng Tong and Chunxiao Jiang},
  journal = {{IEEE} Trans. Veh. Technol.},
  title   = {Diffusion-Based Reinforcement Learning for Cooperative Offloading and Resource Allocation in Multi-{UAV} Assisted Edge-Enabled Metaverse},
  year    = {2025},
  number  = {7},
  pages   = {11281--11293},
  volume  = {74},
}

@Article{Zhao2024a,
  author  = {Yinuo Zhao and Chi Harold Liu and Tianjiao Yi and Guozheng Li and Dapeng Wu},
  journal = {{IEEE} J. Sel. Areas Commun.},
  title   = {Energy-Efficient Ground-Air-Space Vehicular Crowdsensing by Hierarchical Multi-Agent Deep Reinforcement Learning With Diffusion Models},
  year    = {2024},
  number  = {12},
  pages   = {3566--3580},
  volume  = {42},
}

@Article{Zhang2024a,
  author  = {Yiwen Zhang and Zhen Mei and Xiaoqian Wu and Huaiguang Jiang and Jun Zhang and Wenzhong Gao},
  journal = {{IEEE} Trans. Smart Grid},
  title   = {Two-Step Diffusion Policy Deep Reinforcement Learning Method for Low-Carbon Multi-Energy Microgrid Energy Management},
  year    = {2024},
  number  = {5},
  pages   = {4576--4588},
  volume  = {15},
}

@Article{Wang2025b,
  author  = {Wang, Feng and Xie, Jinsong and Zhou, Aimin and Tang, Ke},
  journal = {{IEEE} Trans. Evol. Comput.},
  title   = {A new prediction strategy for dynamic multi-objective optimization using diffusion model},
  year    = {2025},
}

@Article{Fang2024,
  author  = {Fang, Wenxuan and Du, Wei and He, Renchu and Tang, Yang and Jin, Yaochu and Yen, Gary G},
  journal = {{IEEE} Comput. Intell. Mag.},
  title   = {Diffusion model-based multiobjective optimization for gasoline blending scheduling},
  year    = {2024},
  number  = {2},
  pages   = {61--76},
  volume  = {19},
}

@InProceedings{Yang2025,
  author    = {Tzu{-}Yi Yang and Alexander I{-}Chi Lai},
  booktitle = {Proc. {IEEE WCNC}},
  title     = {{WIFIND:} Enhancing {Wi-Fi} Fingerprint Indoor Localization with a Spatially Conditioned Diffusion Model-Based Data Augmentation},
  year      = {2025},
  pages     = {1--6},
}

@Article{Xiang2023,
  author        = {Weilai Xiang and Hongyu Yang and Di Huang and Yunhong Wang},
  journal       = {CoRR},
  title         = {Denoising Diffusion Autoencoders are Unified Self-supervised Learners},
  year          = {2023},
  volume        = {abs/2303.09769},
  archiveprefix = {arXiv},
}

@Article{Chen2024,
  author        = {Xinlei Chen and Zhuang Liu and Saining Xie and Kaiming He},
  journal       = {CoRR},
  title         = {Deconstructing Denoising Diffusion Models for Self-Supervised Learning},
  year          = {2024},
  volume        = {abs/2401.14404},
  archiveprefix = {arXiv},
}

@Article{Xiang2025,
  author        = {Weilai Xiang and Hongyu Yang and Di Huang and Yunhong Wang},
  journal       = {CoRR},
  title         = {{DDAE++:} Enhancing Diffusion Models Towards Unified Generative and Discriminative Learning},
  year          = {2025},
  volume        = {abs/2505.10999},
  archiveprefix = {arXiv},
}

@Article{Sadia2024,
  author        = {Rabeya Tus Sadia and Jie Zhang and Jin Chen},
  journal       = {CoRR},
  title         = {Multiscale Latent Diffusion Model for Enhanced Feature Extraction from Medical Images},
  year          = {2024},
  volume        = {abs/2410.04000},
  archiveprefix = {arXiv},
}

@InProceedings{Hao2024,
  author    = {Shaozhe Hao and Kai Han and Zhengyao Lv and Shihao Zhao and Kwan{-}Yee K. Wong},
  booktitle = {Proc. {ECCV}},
  title     = {{ConceptExpress}: Harnessing Diffusion Models for Single-Image Unsupervised Concept Extraction},
  year      = {2024},
  pages     = {215--233},
  volume    = {15117},
}

@Article{Zhao2024b,
  author        = {Zhao, Changyuan and Wang, Jiacheng and Zhang, Ruichen and Niyato, Dusit and Kim, Dong In and Du, Hongyang},
  journal       = {CoRR},
  title         = {Signal Detection in Near-field Communication with Unknown Noise Characteristics: A Diffusion Model Method},
  year          = {2024},
  volume        = {abs/2409.14031},
  archiveprefix = {arXiv},
}

@Article{Zhang2024b,
  author  = {Ruichen Zhang and Hongyang Du and Yinqiu Liu and Dusit Niyato and Jiawen Kang and Sumei Sun and Xuemin Shen and H. Vincent Poor},
  journal = {{IEEE} Netw.},
  title   = {Interactive {AI} With Retrieval-Augmented Generation for Next Generation Networking},
  year    = {2024},
  number  = {6},
  pages   = {414--424},
  volume  = {38},
}

@Article{Du2025,
  author    = {Du, Hongyang and Zhang, Ruichen and Niyato, Dusit and Kang, Jiawen and Xiong, Zehui and Cui, Shuguang and Shen, Xuemin and Kim, Dong In},
  journal   = {{IEEE} Trans. Pattern Anal. Mach. Intell.},
  title     = {Reinforcement Learning With {LLMs} Interaction For Distributed Diffusion Model Services},
  year      = {2025},
  publisher = {IEEE},
}

@Article{Jiang2024,
  author        = {Shuo Jiang and Min Xie and Jianxi Luo},
  journal       = {CoRR},
  title         = {Large Language Models for Combinatorial Optimization of Design Structure Matrix},
  year          = {2024},
  volume        = {abs/2411.12571},
  archiveprefix = {arXiv},
}

@Article{Li2025a,
  author        = {Jiahui Li and Geng Sun and Zemin Sun and Jiacheng Wang and Yinqiu Liu and Ruichen Zhang and Dusit Niyato and Shiwen Mao},
  journal       = {CoRR},
  title         = {{LLM}-guided {DRL} for Multi-tier {LEO} Satellite Networks with Hybrid {FSO/RF} Links},
  year          = {2025},
  volume        = {abs/2505.11978},
  archiveprefix = {arXiv},
}

@InProceedings{Li2024,
  author    = {Jiawei Li and Yizhe Yang and Yu Bai and Xiaofeng Zhou and Yinghao Li and Huashan Sun and Yuhang Liu and Xingpeng Si and Yuhao Ye and Yixiao Wu and Yiguan Lin and Bin Xu and Ren Bowen and Chong Feng and Yang Gao and Heyan Huang},
  booktitle = {Proc. {ACL}},
  title     = {Fundamental Capabilities of Large Language Models and their Applications in Domain Scenarios: A Survey},
  year      = {2024},
  pages     = {11116--11141},
}

@Article{Sun2025,
  author  = {Jiankai Sun and Chuanyang Zheng and Enze Xie and Zhengying Liu and Ruihang Chu and Jianing Qiu and Jiaqi Xu and Mingyu Ding and Hongyang Li and Mengzhe Geng and Yue Wu and Wenhai Wang and Junsong Chen and Zhangyue Yin and Xiaozhe Ren and Jie Fu and Junxian He and Wu Yuan and Qi Liu and Xihui Liu and Yu Li and Hao Dong and Yu Cheng and Ming Zhang and Pheng{-}Ann Heng and Jifeng Dai and Ping Luo and Jingdong Wang and Ji{-}Rong Wen and Xipeng Qiu and Yike Guo and Hui Xiong and Qun Liu and Zhenguo Li},
  journal = {{ACM} Comput. Surv.},
  title   = {A Survey of Reasoning with Foundation Models: Concepts, Methodologies, and Outlook},
  year    = {2025},
  number  = {11},
  pages   = {278:1--278:43},
  volume  = {57},
}

@InProceedings{Miyake2023,
  author    = {Tamon Miyake and Yushi Wang and Pin{-}Chu Yang and Shigeki Sugano},
  booktitle = {Proc. {ICSR}},
  title     = {Feasibility Study on Parameter Adjustment for a Humanoid Using {LLM} Tailoring Physical Care},
  year      = {2023},
  pages     = {230--243},
  volume    = {14453},
}

@InProceedings{Chen2025,
  author    = {Chen, Bei-Ning and Wei, Feng-Feng and Chen, Wei-Neng},
  booktitle = {Proc. {IEEE ICACI}},
  title     = {Evolutionary Reinforcement Learning with {LLM}-Based Hyperparameter Adaptation},
  year      = {2025},
  pages     = {87--94},
}

@Article{Yan2025,
  author  = {Zijiang Yan and Hao Zhou and Hina Tabassum and Xue Liu},
  journal = {{IEEE} Wirel. Commun. Lett.},
  title   = {Hybrid {LLM-DDQN}-Based Joint Optimization of {V2I} Communication and Autonomous Driving},
  year    = {2025},
  number  = {4},
  pages   = {1214--1218},
  volume  = {14},
}

@Article{Zhou2022,
  author  = {Xiangbing Zhou and Hongjiang Ma and Jianggang Gu and Huiling Chen and Wu Deng},
  journal = {Eng. Appl. Artif. Intell.},
  title   = {Parameter adaptation-based ant colony optimization with dynamic hybrid mechanism},
  year    = {2022},
  pages   = {105139},
  volume  = {114},
}

@Article{Sun2021,
  author  = {Jianyong Sun and Xin Liu and Thomas B{\"{a}}ck and Zongben Xu},
  journal = {{IEEE} Trans. Evol. Comput.},
  title   = {Learning Adaptive Differential Evolution Algorithm From Optimization Experiences by Policy Gradient},
  year    = {2021},
  number  = {4},
  pages   = {666--680},
  volume  = {25},
}

@Article{Chen2024a,
  author  = {Bozhen Chen and Haibin Ouyang and Steven Li and Dexuan Zou},
  journal = {Inf. Sci.},
  title   = {Differential evolution algorithm with a complementary mutation strategy and data Fusion-Based parameter adaptation},
  year    = {2024},
  pages   = {120522},
  volume  = {668},
}

@Article{Zhang2023,
  author        = {Michael R. Zhang and Nishkrit Desai and Juhan Bae and Jonathan Lorraine and Jimmy Ba},
  journal       = {CoRR},
  title         = {Using Large Language Models for Hyperparameter Optimization},
  year          = {2023},
  volume        = {abs/2312.04528},
  archiveprefix = {arXiv},
}

@InProceedings{Li2024a,
  author    = {Hao Li and Xue Yang and Zhaokai Wang and Xizhou Zhu and Jie Zhou and Yu Qiao and Xiaogang Wang and Hongsheng Li and Lewei Lu and Jifeng Dai},
  booktitle = {Proc. {IEEE/CVF CVPR}},
  title     = {Auto {MC-Reward}: Automated Dense Reward Design with Large Language Models for Minecraft},
  year      = {2024},
  pages     = {16426--16435},
}

@InProceedings{Liu2025b,
  author    = {Siyi Liu and Chen Gao and Yong Li},
  booktitle = {Proc. {CPAL}},
  title     = {{AgentHPO}: Large Language Model Agent for Hyper-Parameter Optimization},
  year      = {2025},
  pages     = {1146--1169},
  volume    = {280},
}

@InProceedings{Custode2024,
  author    = {Leonardo Lucio Custode and Fabio Caraffini and Anil Yaman and Giovanni Iacca},
  booktitle = {Proc. {GECCO}},
  title     = {An investigation on the use of Large Language Models for hyperparameter tuning in Evolutionary Algorithms},
  year      = {2024},
  pages     = {1838--1845},
}

@Article{Kochnev2025,
  author        = {Roman Kochnev and Arash Torabi Goodarzi and Zofia Antonina Bentyn and Dmitry Ignatov and Radu Timofte},
  journal       = {CoRR},
  title         = {Optuna vs Code {LlaMa}: Are {LLMs} a New Paradigm for Hyperparameter Tuning?},
  year          = {2025},
  volume        = {abs/2504.06006},
  archiveprefix = {arXiv},
}

@Article{Phan2020,
  author  = {Han Duy Phan and Kirsten Ellis and Jan Carlo Barca and Alan Dorin},
  journal = {Neural Comput. Appl.},
  title   = {A survey of dynamic parameter setting methods for nature-inspired swarm intelligence algorithms},
  year    = {2020},
  number  = {2},
  pages   = {567--588},
  volume  = {32},
}

@Article{Mahammadli2024,
  author        = {Kanan Mahammadli},
  journal       = {CoRR},
  title         = {Sequential Large Language Model-Based Hyper-Parameter Optimization},
  year          = {2024},
  volume        = {abs/2410.20302},
  archiveprefix = {arXiv},
}

@Article{Sun2025a,
  author  = {Geng Sun and Yixian Wang and Dusit Niyato and Jiacheng Wang and Xinying Wang and H. Vincent Poor and Khaled B. Letaief},
  journal = {{IEEE} Netw.},
  title   = {Large Language Model {(LLM)}-Enabled Graphs in Dynamic Networking},
  year    = {2025},
  number  = {4},
  pages   = {290--301},
  volume  = {39},
}

@Article{Ghosh2022,
  author  = {Arka Ghosh and Swagatam Das and Asit Kumar Das and Roman Senkerik and Adam Viktorin and Ivan Zelinka and Antonio David Masegosa},
  journal = {Swarm Evol. Comput.},
  title   = {Using spatial neighborhoods for parameter adaptation: An improved success history based differential evolution},
  year    = {2022},
  pages   = {101057},
  volume  = {71},
}

@InProceedings{Tanabe2013,
  author    = {Ryoji Tanabe and Alex Fukunaga},
  booktitle = {Proc. {IEEE CEC}},
  title     = {Success-history based parameter adaptation for Differential Evolution},
  year      = {2013},
  pages     = {71--78},
}

@Article{Viktorin2019,
  author  = {Adam Viktorin and Roman Senkerik and Michal Pluhacek and Tomas Kadavy and Ales Zamuda},
  journal = {Swarm Evol. Comput.},
  title   = {Distance based parameter adaptation for Success-History based Differential Evolution},
  year    = {2019},
  volume  = {50},
}

@Article{Melin2013,
  author  = {Patricia Melin and Frumen Olivas and Oscar Castillo and Fevrier Valdez and Jose Soria and Jos{\'{e}} Mario Garc{\'{\i}}a Valdez},
  journal = {Expert Syst. Appl.},
  title   = {Optimal design of fuzzy classification systems using {PSO} with dynamic parameter adaptation through fuzzy logic},
  year    = {2013},
  number  = {8},
  pages   = {3196--3206},
  volume  = {40},
}

@Article{TranDang2020,
  author  = {Hoa Tran{-}Dang and Nicolas Krommenacker and Patrick Charpentier and Dong{-}Seong Kim},
  journal = {{IEEE} Internet Things J.},
  title   = {Toward the Internet of Things for Physical Internet: Perspectives and Challenges},
  year    = {2020},
  number  = {6},
  pages   = {4711--4736},
  volume  = {7},
}

@Article{Alao2025,
  author    = {Alao, Damilola and Younang, Victorine Clotilde Wakam and Sen, Amartya},
  journal   = {{IEEE} Trans. Serv. Comput.},
  title     = {Extending {IoT} Devices Network Lifetime Using {IoT} Service Prediction and Recommendation},
  year      = {2025},
  publisher = {IEEE},
}

@Article{Sheng2025,
  author  = {Chuan Sheng and Wei Zhou and Qing{-}Long Han and Wanlun Ma and Xiaogang Zhu and Sheng Wen and Yang Xiang},
  journal = {{IEEE} Trans. Ind. Informatics},
  title   = {Network Traffic Fingerprinting for {IIoT} Device Identification: A Survey},
  year    = {2025},
  number  = {5},
  pages   = {3541--3554},
  volume  = {21},
}

@Article{Dong2021a,
  author  = {Shi Dong and Yuanjun Xia and Tao Peng},
  journal = {{IEEE} Trans. Netw. Serv. Manag.},
  title   = {Network Abnormal Traffic Detection Model Based on Semi-Supervised Deep Reinforcement Learning},
  year    = {2021},
  number  = {4},
  pages   = {4197--4212},
  volume  = {18},
}

@Article{Liu2024,
  author  = {Ya Liu and Xiao Wang and Bo Qu and Fengyu Zhao},
  journal = {{IEEE} Trans. Inf. Forensics Secur.},
  title   = {{ATVITSC:} A Novel Encrypted Traffic Classification Method Based on Deep Learning},
  year    = {2024},
  pages   = {9374--9389},
  volume  = {19},
}

@Article{Wu2021,
  author  = {Hao Wu and Xi Zhang and Jufeng Yang},
  journal = {J. Web Eng.},
  title   = {Deep Learning-Based Encrypted Network Traffic Classification and Resource Allocation in {SDN}},
  year    = {2021},
  number  = {8},
  volume  = {20},
}

@Article{Dai2023,
  author  = {Jianbang Dai and Xiaolong Xu and Fu Xiao},
  journal = {Comput. Networks},
  title   = {{GLADS:} A global-local attention data selection model for multimodal multitask encrypted traffic classification of {IoT}},
  year    = {2023},
  pages   = {109652},
  volume  = {225},
}

@Article{Xu2024,
  author  = {Xu, Junzhi and Zhang, Yibin and Zhou, Kaijie and Wang, Qin and Hua, Minyu and Shan, Lin and Lin, Yun and Gui, Guan},
  journal = {{IEEE} Trans. Cogn. Commun. Netw.},
  title   = {A Cascaded Broad Learning Network Embedded Image Features for Malware Traffic Classification},
  year    = {2024},
}

@Article{Rezaei2019,
  author  = {Shahbaz Rezaei and Xin Liu},
  journal = {{IEEE} Commun. Mag.},
  title   = {Deep Learning for Encrypted Traffic Classification: An Overview},
  year    = {2019},
  number  = {5},
  pages   = {76--81},
  volume  = {57},
}

@InProceedings{Zhou2017,
  author    = {Huiyi Zhou and Yong Wang and Xiaochun Lei and Yuming Liu},
  booktitle = {Proc. {CIS}},
  title     = {A Method of Improved {CNN} Traffic Classification},
  year      = {2017},
  pages     = {177--181},
}

@Article{Xiao2022,
  author  = {Xi Xiao and Wentao Xiao and Rui Li and Xiapu Luo and Haitao Zheng and Shutao Xia},
  journal = {{IEEE} Trans. Dependable Secur. Comput.},
  title   = {{EBSNN:} Extended Byte Segment Neural Network for Network Traffic Classification},
  year    = {2022},
  number  = {5},
  pages   = {3521--3538},
  volume  = {19},
}

@Article{Zhu2023,
  author  = {Shizhou Zhu and Xiaolong Xu and Honghao Gao and Fu Xiao},
  journal = {{IEEE} Internet Things J.},
  title   = {{CMTSNN:} A Deep Learning Model for Multiclassification of Abnormal and Encrypted Traffic of Internet of Things},
  year    = {2023},
  number  = {13},
  pages   = {11773--11791},
  volume  = {10},
}

@InProceedings{Ede2020,
  author    = {Thijs van Ede and Riccardo Bortolameotti and Andrea Continella and Jingjing Ren and Daniel J. Dubois and Martina Lindorfer and David R. Choffnes and Maarten van Steen and Andreas Peter},
  booktitle = {Proc. {NDSS}},
  title     = {{FlowPrint}: Semi-Supervised Mobile-{App} Fingerprinting on Encrypted Network Traffic},
  year      = {2020},
}

@Article{Lin2021,
  author  = {Kunda Lin and Xiaolong Xu and Honghao Gao},
  journal = {Comput. Networks},
  title   = {{TSCRNN:} A novel classification scheme of encrypted traffic based on flow spatiotemporal features for efficient management of {IIoT}},
  year    = {2021},
  pages   = {107974},
  volume  = {190},
}

@InProceedings{Zhang2025c,
  author    = {Haozhen Zhang and Haodong Yue and Xi Xiao and Le Yu and Qing Li and Zhen Ling and Ye Zhang},
  booktitle = {Proc. {AAAI}},
  title     = {Revolutionizing Encrypted Traffic Classification with {MH-Net}: A Multi-View Heterogeneous Graph Model},
  year      = {2025},
  pages     = {1048--1056},
}

@Article{Zhang2019,
  author  = {Chaoyun Zhang and Paul Patras and Hamed Haddadi},
  journal = {{IEEE} Commun. Surv. Tutorials},
  title   = {Deep Learning in Mobile and Wireless Networking: A Survey},
  year    = {2019},
  number  = {3},
  pages   = {2224--2287},
  volume  = {21},
}

@Article{Wang2021,
  author  = {Pan Wang and Zixuan Wang and Feng Ye and Xuejiao Chen},
  journal = {Comput. Networks},
  title   = {{ByteSGAN}: A semi-supervised Generative Adversarial Network for encrypted traffic classification in {SDN} Edge Gateway},
  year    = {2021},
  pages   = {108535},
  volume  = {200},
}

@Article{Tong2024,
  author  = {Jianheng Tong and Ying Zhang},
  journal = {{IEEE} Internet Things J.},
  title   = {A Real-Time Label-Free Self-Supervised Deep Learning Intrusion Detection for Handling New Type and Few-Shot Attacks in {IoT} Networks},
  year    = {2024},
  number  = {19},
  pages   = {30769--30786},
  volume  = {11},
}

@Article{Nakip2024,
  author  = {Mert Nakip and Erol Gelenbe},
  journal = {{IEEE} Trans. Inf. Forensics Secur.},
  title   = {Online Self-Supervised Deep Learning for Intrusion Detection Systems},
  year    = {2024},
  pages   = {5668--5683},
  volume  = {19},
}

@Article{De2022,
  author  = {De, Suparna and Bermudez-Edo, Maria and Xu, Honghui and Cai, Zhipeng},
  journal = {{IEEE} Trans. Ind. Informatics},
  title   = {Deep generative models in the industrial internet of things: A survey},
  year    = {2022},
  number  = {9},
  pages   = {5728--5737},
  volume  = {18},
}

@Article{Abbasi2021,
  author  = {Mahmoud Abbasi and Amin Shahraki and Amir Taherkordi},
  journal = {Comput. Commun.},
  title   = {Deep Learning for Network Traffic Monitoring and Analysis {(NTMA):} A Survey},
  year    = {2021},
  pages   = {19--41},
  volume  = {170},
}

@Article{Song2024,
  author  = {Xianfang Song and Yong Zhang and Wanqiu Zhang and Chunlin He and Ying Hu and Jian Wang and Dunwei Gong},
  journal = {Swarm Evol. Comput.},
  title   = {Evolutionary computation for feature selection in classification: A comprehensive survey of solutions, applications and challenges},
  year    = {2024},
  pages   = {101661},
  volume  = {90},
}

@Article{Nguyen2012,
  author  = {Thuy T. T. Nguyen and Grenville J. Armitage and Philip Branch and Sebastian Zander},
  journal = {{IEEE/ACM} Trans. Netw.},
  title   = {Timely and continuous machine-learning-based classification for interactive {IP} traffic},
  year    = {2012},
  number  = {6},
  pages   = {1880--1894},
  volume  = {20},
}

@Article{Gaurav2023,
  author  = {Akshat Gaurav and Brij B. Gupta and Prabin Kumar Panigrahi},
  journal = {Enterp. Inf. Syst.},
  title   = {A comprehensive survey on machine learning approaches for malware detection in {IoT}-based enterprise information system},
  year    = {2023},
  number  = {3},
  volume  = {17},
}

@InProceedings{Yan2019,
  author    = {Haonan Yan and Hui Li and Mingchi Xiao and Rui Dai and Xianchun Zheng and Xingwen Zhao and Fenghua Li},
  booktitle = {Proc. {IEEE GLOBECOM}},
  title     = {{PGSM-DPI:} Precisely Guided Signature Matching of Deep Packet Inspection for Traffic Analysis},
  year      = {2019},
  pages     = {1--6},
}

@Article{Blaise2020,
  author  = {Agathe Blaise and Mathieu Bouet and Vania Conan and Stefano Secci},
  journal = {Comput. Networks},
  title   = {Detection of zero-day attacks: An unsupervised port-based approach},
  year    = {2020},
  pages   = {107391},
  volume  = {180},
}

@Article{Papadogiannaki2022,
  author  = {Eva Papadogiannaki and Sotiris Ioannidis},
  journal = {{ACM} Comput. Surv.},
  title   = {A Survey on Encrypted Network Traffic Analysis Applications, Techniques, and Countermeasures},
  year    = {2022},
  number  = {6},
  pages   = {123:1--123:35},
  volume  = {54},
}

@Article{Nascita2024,
  author    = {Nascita, Alfredo and Aceto, Giuseppe and Ciuonzo, Domenico and Montieri, Antonio and Persico, Valerio and Pescap{\'e}, Antonio},
  journal   = {{IEEE} Commun. Surv. Tutorials},
  title     = {A survey on explainable artificial intelligence for internet traffic classification and prediction, and intrusion detection},
  year      = {2024},
  publisher = {IEEE},
}

@Article{Zhang2015,
  author  = {Jun Zhang and Xiao Chen and Yang Xiang and Wanlei Zhou and Jie Wu},
  journal = {{IEEE/ACM} Trans. Netw.},
  title   = {Robust Network Traffic Classification},
  year    = {2015},
  number  = {4},
  pages   = {1257--1270},
  volume  = {23},
}

@Article{Wei2022,
  author  = {Wenting Wei and Huaxi Gu and Wenshuai Deng and Zhe Xiao and Xinming Ren},
  journal = {Neurocomputing},
  title   = {{ABL-TC:} A lightweight design for network traffic classification empowered by deep learning},
  year    = {2022},
  pages   = {333--344},
  volume  = {489},
}

@Article{Liu2024a,
  author  = {Guangyuan Liu and Hongyang Du and Dusit Niyato and Jiawen Kang and Zehui Xiong and Dong In Kim and Xuemin Shen},
  journal = {{IEEE} Netw.},
  title   = {Semantic Communications for Artificial Intelligence Generated Content {(AIGC)} Toward Effective Content Creation},
  year    = {2024},
  number  = {5},
  pages   = {295--303},
  volume  = {38},
}

@Article{Li2024b,
  author  = {Zhiyuan Li and Xiaoping Xu},
  journal = {Comput. Networks},
  title   = {{L2-BiTCN-CNN}: Spatio-temporal features fusion-based multi-classification model for various internet applications identification},
  year    = {2024},
  pages   = {110298},
  volume  = {243},
}

@Article{Ho2020,
  author  = {Ho, Jonathan and Jain, Ajay and Abbeel, Pieter},
  journal = {Advances in neural information processing systems},
  title   = {Denoising diffusion probabilistic models},
  year    = {2020},
  pages   = {6840--6851},
  volume  = {33},
}

@Article{Abbasi2022,
  author  = {Muhammad Shabbir Abbasi and Harith Al{-}Sahaf and Masood Mansoori and Ian Welch},
  journal = {Appl. Soft Comput.},
  title   = {Behavior-based ransomware classification: A particle swarm optimization wrapper-based approach for feature selection},
  year    = {2022},
  pages   = {108744},
  volume  = {121},
}

@InProceedings{DraperGil2016,
  author    = {Gerard Draper{-}Gil and Arash Habibi Lashkari and Mohammad Saiful Islam Mamun and Ali A. Ghorbani},
  booktitle = {Proc. {ICISSP}},
  title     = {Characterization of Encrypted and {VPN} Traffic using Time-related Features},
  year      = {2016},
  pages     = {407--414},
  publisher = {SciTePress},
}

@Article{Ferrag2022,
  author  = {Mohamed Amine Ferrag and Othmane Friha and Djallel Hamouda and Leandros Maglaras and Helge Janicke},
  journal = {{IEEE} Access},
  title   = {Edge-IIoTset: A New Comprehensive Realistic Cyber Security Dataset of IoT and IIoT Applications for Centralized and Federated Learning},
  year    = {2022},
  pages   = {40281--40306},
  volume  = {10},
}

@InProceedings{Aceto2018,
  author    = {Giuseppe Aceto and Domenico Ciuonzo and Antonio Montieri and Antonio Pescap{\`{e}}},
  booktitle = {Proc. {IEEE TMA}},
  title     = {Mobile Encrypted Traffic Classification Using Deep Learning},
  year      = {2018},
  pages     = {1--8},
}

@Article{Liu2021,
  author  = {Dunnan Liu and Xiaofeng Xu and Mingguang Liu and Yaling Liu},
  journal = {Neural Comput. Appl.},
  title   = {Dynamic traffic classification algorithm and simulation of energy Internet of things based on machine learning},
  year    = {2021},
  number  = {9},
  pages   = {3967--3976},
  volume  = {33},
}

@Article{Donato2014,
  author  = {Walter de Donato and Antonio Pescap{\`{e}} and Alberto Dainotti},
  journal = {{IEEE} Netw.},
  title   = {Traffic identification engine: an open platform for traffic classification},
  year    = {2014},
  number  = {2},
  pages   = {56--64},
  volume  = {28},
}

@Article{Dainotti2012,
  author  = {Alberto Dainotti and Antonio Pescap{\`{e}} and Kimberly C. Claffy},
  journal = {{IEEE} Netw.},
  title   = {Issues and future directions in traffic classification},
  year    = {2012},
  number  = {1},
  pages   = {35--40},
  volume  = {26},
}

@InProceedings{Doroud2018,
  author    = {Hossein Doroud and Giuseppe Aceto and Walter de Donato and Elnaz Alizadeh Jarchlo and Andr{\'{e}}s Mar{\'{\i}}n L{\'{o}}pez and C{\'{e}}sar D. Guerrero and Antonio Pescap{\`{e}}},
  booktitle = {Proc. {IEEE GLOBECOM}},
  title     = {Speeding-Up {DPI} Traffic Classification with Chaining},
  year      = {2018},
  pages     = {1--6},
}

@Article{Liang2022,
  author  = {Wei Liang and Yiyong Hu and Xiaokang Zhou and Yi Pan and Kevin I{-}Kai Wang},
  journal = {{IEEE} Trans. Ind. Informatics},
  title   = {Variational Few-Shot Learning for Microservice-Oriented Intrusion Detection in Distributed Industrial {IoT}},
  year    = {2022},
  number  = {8},
  pages   = {5087--5095},
  volume  = {18},
}

@InProceedings{He2020,
  author    = {Kaiming He and Haoqi Fan and Yuxin Wu and Saining Xie and Ross B. Girshick},
  booktitle = {Proc. {CVPR}},
  title     = {Momentum Contrast for Unsupervised Visual Representation Learning},
  year      = {2020},
  pages     = {9726--9735},
}

@Article{Sadeghzadeh2021,
  author  = {Amir Mahdi Sadeghzadeh and Saeed Shiravi and Rasool Jalili},
  journal = {{IEEE} Trans. Netw. Serv. Manag.},
  title   = {Adversarial Network Traffic: Towards Evaluating the Robustness of Deep-Learning-Based Network Traffic Classification},
  year    = {2021},
  number  = {2},
  pages   = {1962--1976},
  volume  = {18},
}

@Article{Xiong2014,
  author    = {Wei Xiong and Hanping Hu and Naixue Xiong and Laurence T. Yang and Wen{-}Chih Peng and Xiaofei Wang and Yanzhen Qu},
  journal   = {Inf. Sci.},
  title     = {Anomaly secure detection methods by analyzing dynamic characteristics of the network traffic in cloud communications},
  year      = {2014},
  pages     = {403--415},
  volume    = {258},
}

@Article{Zhang2013,
  author  = {Jun Zhang and Yang Xiang and Yu Wang and Wanlei Zhou and Yong Xiang and Yong Guan},
  journal = {{IEEE} Trans. Parallel Distributed Syst.},
  title   = {Network Traffic Classification Using Correlation Information},
  year    = {2013},
  number  = {1},
  pages   = {104--117},
  volume  = {24},
}

@Article{Ren2021,
  author    = {Xinming Ren and Huaxi Gu and Wenting Wei},
  journal   = {Expert Syst. Appl.},
  title     = {{Tree-RNN}: Tree structural recurrent neural network for network traffic classification},
  year      = {2021},
  pages     = {114363},
  volume    = {167},
}

@Article{Booij2022,
  author  = {Tim M. Booij and Irina Chiscop and Erik Meeuwissen and Nour Moustafa and Frank T. H. den Hartog},
  journal = {{IEEE} Internet Things J.},
  title   = {{ToN\_IoT}: The Role of Heterogeneity and the Need for Standardization of Features and Attack Types in {IoT} Network Intrusion Data Sets},
  year    = {2022},
  number  = {1},
  pages   = {485--496},
  volume  = {9},
}

@Article{Ding2022,
  author  = {Yi Ding and Guiqin Zhu and Dajiang Chen and Xue Qin and Mingsheng Cao and Zhiguang Qin},
  journal = {{IEEE} Trans. Intell. Transp. Syst.},
  title   = {Adversarial Sample Attack and Defense Method for Encrypted Traffic Data},
  year    = {2022},
  number  = {10},
  pages   = {18024--18039},
  volume  = {23},
}

@InProceedings{Zhang2020,
  author    = {Jielun Zhang and Fuhao Li and Feng Ye and Hongyu Wu},
  booktitle = {Proc. {IEEE INFOCOM}},
  title     = {Autonomous Unknown-Application Filtering and Labeling for {DL}-based Traffic Classifier Update},
  year      = {2020},
  pages     = {397--405},
}

@Article{He2025,
  author  = {Chunming He and Yuqi Shen and Chengyu Fang and Fengyang Xiao and Longxiang Tang and Yulun Zhang and Wangmeng Zuo and Zhenhua Guo and Xiu Li},
  journal = {{IEEE} Trans. Pattern Anal. Mach. Intell.},
  title   = {Diffusion Models in Low-Level Vision: A Survey},
  year    = {2025},
  number  = {6},
  pages   = {4630--4651},
  volume  = {47},
}

@InProceedings{Shi2010,
  author    = {Na Shi and Xumin Liu and Yong Guan},
  booktitle = {Proc. {IITSI}},
  title     = {Research on k-means Clustering Algorithm: An Improved k-means Clustering Algorithm},
  year      = {2010},
  pages     = {63--67},
}

@Article{Yang2003,
  author  = {Jian Yang and Jing{-}Yu Yang and David Zhang and Jianfeng Lu},
  journal = {Pattern Recognit.},
  title   = {Feature fusion: parallel strategy vs. serial strategy},
  year    = {2003},
  number  = {6},
  pages   = {1369--1381},
  volume  = {36},
}

@Article{Diao2015,
  author    = {Ren Diao and Qiang Shen},
  journal   = {Artif. Intell. Rev.},
  title     = {Nature inspired feature selection meta-heuristics},
  year      = {2015},
  number    = {3},
  pages     = {311--340},
  volume    = {44},
}

@Article{Croitoru2023,
  author  = {Florinel{-}Alin Croitoru and Vlad Hondru and Radu Tudor Ionescu and Mubarak Shah},
  journal = {{IEEE} Trans. Pattern Anal. Mach. Intell.},
  title   = {Diffusion Models in Vision: A Survey},
  year    = {2023},
  number  = {9},
  pages   = {10850--10869},
  volume  = {45},
}

@InProceedings{Ginige2024,
  author    = {Ginige, Yasod and Dahanayaka, Thilini and Seneviratne, Suranga},
  booktitle = {Proc. {AINTEC}},
  title     = {{TrafficGPT}: An {LLM} Approach for Open-Set Encrypted Traffic Classification},
  year      = {2024},
  pages     = {26--35},
}

@InProceedings{zhou2024enhancing,
  author       = {Zhou, Honghe and Huang, Xin and Deng, Lin},
  booktitle    = {Proc .{BigData}},
  title        = {Enhancing Network Traffic Classification with Large Language Models},
  year         = {2024},
  organization = {IEEE},
  pages        = {7282--7291},
}

@Article{chen2024merlot,
  author  = {Chen, Yuxuan and Li, Rongpeng and Zhao, Zhifeng and Zhang, Honggang},
  journal = {arXiv preprint arXiv:2411.13004},
  title   = {{MERLOT}: A Distilled {LLM}-based Mixture-of-Experts Framework for Scalable Encrypted Traffic Classification},
  year    = {2024},
}

@Article{Xie2025,
  author  = {Xie, Wenwen and Sun, Geng and Li, Jiahui and Wang, Jiacheng and Du, Hongyang and Niyato, Dusit and Dobre, Octavia A},
  journal = {{IEEE} Internet Things Mag.},
  title   = {Generative {AI} for Energy Harvesting Internet of Things Network: Fundamental, Applications, and Opportunities},
  year    = {2025},
  number  = {3},
  pages   = {72--80},
  volume  = {8},
}

@Article{Stellios2018,
  author  = {Ioannis Stellios and Panayiotis Kotzanikolaou and Mihalis Psarakis and Cristina Alcaraz and Javier L{\'{o}}pez},
  journal = {{IEEE} Commun. Surv. Tutorials},
  title   = {A Survey of {IoT}-Enabled Cyberattacks: Assessing Attack Paths to Critical Infrastructures and Services},
  year    = {2018},
  number  = {4},
  pages   = {3453--3495},
  volume  = {20},
}

@Article{Sun2025b,
  author  = {Geng Sun and Wenwen Xie and Dusit Niyato and Fang Mei and Jiawen Kang and Hongyang Du and Shiwen Mao},
  journal = {{IEEE} Wirel. Commun.},
  title   = {Generative {AI} for Deep Reinforcement Learning: Framework, Analysis, and Use Cases},
  year    = {2025},
  number  = {3},
  pages   = {186--195},
  volume  = {32},
}

@Article{Zhang2025d,
  author  = {Chuang Zhang and Geng Sun and Jiahui Li and Qingqing Wu and Jiacheng Wang and Dusit Niyato and Yuanwei Liu},
  journal = {{IEEE} Trans. Mob. Comput.},
  title   = {Multi-Objective Aerial Collaborative Secure Communication Optimization via Generative Diffusion Model-Enabled Deep Reinforcement Learning},
  year    = {2025},
  number  = {4},
  pages   = {3041--3058},
  volume  = {24},
}

@Article{Sun2025d,
  author    = {Geng Sun and Wenwen Xie and Dusit Niyato and Hongyang Du and Jiawen Kang and Jing Wu and Sumei Sun and Ping Zhang},
  journal   = {{IEEE} Netw.},
  title     = {Generative {AI} for Advanced {UAV} Networking},
  year      = {2025},
  number    = {4},
  pages     = {244--253},
  volume    = {39},
}

@Article{Zheng2025a,
  author        = {Xiaoya Zheng and Geng Sun and Jiahui Li and Jiacheng Wang and Qingqing Wu and Dusit Niyato and Abbas Jamalipour},
  journal       = {CoRR},
  title         = {{UAV} Swarm-enabled Collaborative Post-disaster Communications in Low Altitude Economy via a Two-stage Optimization Approach},
  year          = {2025},
  volume        = {abs/2501.05742},
  archiveprefix = {arXiv},
}

@Article{Li2024c,
  author  = {Jiahui Li and Geng Sun and Lingjie Duan and Qingqing Wu},
  journal = {{IEEE} Trans. Mob. Comput.},
  title   = {Multi-Objective Optimization for {UAV} Swarm-Assisted {IoT} With Virtual Antenna Arrays},
  year    = {2024},
  number  = {5},
  pages   = {4890--4907},
  volume  = {23},
}

@ARTICLE{OBETrans,
author={Wang, Zhe and Zhang, Jiayi and Bj{\"o}rnson, Emil and Niyato, Dusit and Ai, Bo},
title = {Optimal Bilinear Equalizer for Cell-Free Massive {MIMO} Systems over Correlated {R}ician Channels},
 journal={IEEE Trans. Signal Process.},
year = {2025}}

@Article{Zhang2014a,
  author    = {Yudong Zhang and Shuihua Wang and Preetha Phillips and Genlin Ji},
  journal   = {Knowl. Based Syst.},
  title     = {Binary {PSO} with mutation operator for feature selection using decision tree applied to spam detection},
  year      = {2014},
  pages     = {22--31},
  volume    = {64},
}

@Article{Li2025c,
  author  = {Hongjuan Li and Haiyuan Chen and Miao Wang and Jiahui Li and Hui Kang and Yuzhuo Guan and Xu Lin},
  journal = {Eng. Appl. Artif. Intell.},
  title   = {Multi-objective deployment optimization for integrated sensing and communication-enabled unmanned aerial vehicle swarm},
  year    = {2025},
  pages   = {112368},
  volume  = {162},
}

@Article{Clerc2002,
  author    = {Maurice Clerc and James Kennedy},
  journal   = {{IEEE} Trans. Evol. Comput.},
  title     = {The particle swarm - explosion, stability, and convergence in a multidimensional complex space},
  year      = {2002},
  number    = {1},
  pages     = {58--73},
  volume    = {6},
}

@Article{Trelea2003,
  author    = {Ioan Cristian Trelea},
  journal   = {Inf. Process. Lett.},
  title     = {The particle swarm optimization algorithm: convergence analysis and parameter selection},
  year      = {2003},
  number    = {6},
  pages     = {317--325},
  volume    = {85},
}

@Article{Li2025Aerial,
  author        = {Jiahui Li and Geng Sun and Qingqing Wu and Shuang Liang and Jiacheng Wang and Dusit Niyato and Dong In Kim},
  journal       = {CoRR},
  title         = {Aerial Secure Collaborative Communications under Eavesdropper Collusion in Low-altitude Economy: {A} Generative Swarm Intelligent Approach},
  year          = {2025},
  volume        = {abs/2503.00721},
  archiveprefix = {arXiv},
}

\end{document}